\documentclass[aip, pop, preprint]{revtex4-1}

\usepackage{graphicx}
\usepackage{wrapfig}
\usepackage{amsmath,amssymb}
\usepackage[utf8]{inputenc}
\usepackage[rightcaption]{sidecap}
\usepackage{caption}

\newcommand{\be}[1]{\begin{equation} \label{#1}}
\newcommand{\ee}{\end{equation}}
\newcommand{\bea}[1]{\begin{eqnarray} \label{#1}}
\newcommand{\eea}{\end{eqnarray}}
\newcommand{\bean}{\begin{eqnarray*}}
\newcommand{\eean}{\end{eqnarray*}}

\newcommand{\eq}[1]{(\ref{#1})}
\newcommand{\difp}[2]{\frac{\partial #1}{\partial #2}}
\newcommand{\br}{{\bf r}}

\newcommand{\bA}{{\bf A}}
\newcommand{\ba}{{\bf a}}
\newcommand{\bB}{{\bf B}}
\newcommand{\bb}{{\bf b}}
\newcommand{\bE}{{\bf E}}
\newcommand{\bn}{{\bf n}}

\newcommand{\bp}{{\bf p}}
\newcommand{\bP}{{\bf P}}

\newcommand{\by}{{\bf y}}
\newcommand{\bz}{{\bf z}}

\newcommand{\bx}{{\bf x}}

\newcommand{\rd}{{\rm d}}

\usepackage{xcolor}
\usepackage[normalem]{ulem}

\newcommand{\red}[1]{#1} % add
\newcommand{\blue}[1]{}  % remove
\newcommand{\green}[1]{} % comment

\newcommand{\redx}[1]{#1} % add
\newcommand{\bluex}[1]{}  % remove
\newcommand{\greenx}[1]{} % comment

\begin{document}

\title{\red{Quasi-}geometric integration of guiding-center orbits in piecewise linear toroidal fields}
\author{M. Eder}
\thanks{Author to whom correspondence should be addressed}
\email{eder@tugraz.at}
\affiliation{Fusion@ÖAW, Institut für Theoretische Physik - Computational Physics, Technische Universität Graz, Petersgasse 16, 8010 Graz, Austria}
\author{C.G. Albert}
\affiliation{Max-Planck-Institut für Plasmaphysik, Boltzmannstr. 2, 85748 Garching, Germany}
\author{L.M.P. Bauer}
\affiliation{Fusion@ÖAW, Institut für Theoretische Physik - Computational Physics, Technische Universität Graz, Petersgasse 16, 8010 Graz, Austria}
\author{S.V. Kasilov}
\affiliation{Fusion@ÖAW, Institut für Theoretische Physik - Computational Physics, Technische Universität Graz, Petersgasse 16, 8010 Graz, Austria}
\affiliation{Institute of Plasma Physics, National Science Center ‘‘Kharkov Institute of Physics and Technology’’, \redx{Akademicheskaya str. 1,} 61108, Kharkov, Ukraine}
\affiliation{\redx{V.N.Karazin Kharkov National University, Svobody sq. 4, 61022, Kharkov, Ukraine}}
\author{W. Kernbichler}
\affiliation{Fusion@ÖAW, Institut für Theoretische Physik - Computational Physics, Technische Universität Graz, Petersgasse 16, 8010 Graz, Austria}

\begin{abstract}

A \blue{geometric}\red{numerical} integration method for guiding-center orbits of charged particles in toroidal fusion devices with three-dimensional field geometry is described.
Here, high order interpolation of electromagnetic fields in space is replaced by a special linear interpolation, leading to locally linear Hamiltonian equations of motion with piecewise constant coefficients.
This approach reduces computational effort and noise sensitivity while the conservation of total energy, magnetic moment and phase space volume is retained.
\red{The underlying formulation treats motion in piecewise linear fields exactly and thus preserves the non-canonical symplectic form. The algorithm itself
is only quasi-geometric due to a series expansion in the orbit parameter. For practical
purposes an expansion to the fourth order retains geometric properties down to computer accuracy in typical examples.}
When applied to collisionless guiding-center orbits in an axisymmetric tokamak and a realistic
three-dimensional stellarator configuration, the method demonstrates \bluex{correct}\redx{stable} long-term orbit dynamics
\redx{conserving invariants}.
\blue{Within}\red{In} Monte Carlo evaluation of transport coefficients, the computational efficiency of \red{quasi-}geometric
integration is an order of magnitude higher than with a standard \blue{adaptive}\red{fourth order} Runge-Kutta integrator.\\

\end{abstract}

\maketitle

\clearpage

\section{Introduction}

Global kinetic computations of quasi-steady plasma parameters in 3D toroidal fusion devices utilize the evaluation of the distribution function and/or its moments by direct modeling of particle orbits. This includes Monte Carlo transport simulations in given external fields~\citep{boozer_monte_1981,lotz_monte_1988, heikkinen_particle_2001, wakasa_study_2008, tribaldos_monte_2001, isaev_venusf_2006,allmaier_variance_2008, satake_neoclassical_2011, pfefferle_venus-levis_2014} as well as self-consistent turbulence models with particle codes.~\citep{jolliet_global_2007,mishchenko_global_2008,hager_gyrokinetic_2016,ku_fast_2018} Kinetic modeling of 3D plasma equilibria~\cite{albert_kinetic_2016} or edge plasmas~\cite{hager_gyrokinetic_2019} puts specific requirements on solving the guiding-center equations.~\citep{morozov_structure_1966,boozer_guiding_1980,littlejohn_variational_1983,cary_hamiltonian_2009} Namely, particle orbit integration should be computationally efficient, tolerant to statistical noise in the electromagnetic field and efficient in scoring statistical data from the orbits.
Geometric integrators~\citep{hairer_geometric_2006,morrison_structure_2017} address these targets by relaxing the requirement to the accuracy of guiding-center orbits while preserving physically \bluex{correct}\redx{consistent} long time orbit dynamics. In this context the word geometric refers to the preservation of the geometric structure of phase-space being a symplectic manifold.
The \blue{most well-known}\red{best-known} class of geometric integrators are symplectic integrators that rely on canonical coordinates in phase-space. Symplectic integrators are not directly applicable to guiding-center dynamics that are formulated in non-canonical coordinates. One way to circumvent this problem is the use of a (usually implicit) transformation to canonical coordinates.~\citep{zhang_canonicalization_2014,zhu_symmetric_2016,albert_symplectic_2020} Up to now such approaches rely either on magnetic flux coordinates or require a more expensive transformation of phase-space coordinates in the general case. The other
\red{well-}known alternative are variational integrators.~\citep{qin_variational_2008,qin_variational_2009} Such integrators do not assume canonical coordinates and include symplectic integrators as a special case. Stability problems of variational integrators arise for guiding-center orbit computations due to the degeneracy of their phase-space Lagrangian. This issue has only recently been resolved.~\citep{burby_toroidal_2017,kraus_projected_2017,ellison_degenerate_2018} The resulting integrators are fully implicit and/or require an augmented set of dynamical variables, and their competitiveness for practical
applications is still a topic open to investigation. \red{Yet another alternative
could arise from very recent results on slow-manifold methods~\cite{xiao_slow_2020}
that introduce a different construction of the guiding-center equations well suited for geometric integration.}

The method presented here is geometric in both\blue{,} a structure-preserving and a more literal sense, as it considers
intersections of orbits with a spatial mesh. \red{The underlying formulation and discretization
of fields exactly preserves the non-canonical Hamiltonian structure of the equations.
\redx{In contrast to usual geometric integrators, this is achieved by solving
the exact motion in simplified fields, i.e. the Hamiltonian flow in the
original fields is approximated by the exact Hamiltonian flow in piecewise
linear fields.}
The final algorithm is only quasi-geometric, as it relies on a series expansion in the orbit parameter.
\redx{Here the term quasi-geometric means that the error can be brought below
any given (computer) accuracy by a sufficiently high order.}
\bluex{but}\redx{Expansions to order 3 and 4 are} \bluex{is} shown to conserve invariants extremely well for at least $10^6$ toroidal turns in numerical experiments for typical fusion devices.}
\blue{It}\red{The approach} has been introduced in Ref.~\onlinecite{eder_three-dimensional_2019}
as a generalization of the 2D geometric integrator of Ref.~\onlinecite{kasilov_geometric_2016} for general 3D toroidal fields.
This approach has two features useful for application in global equilibrium and transport simulations: straightforward computation of
spatial distributions of macroscopic parameters and robustness
in the presence of noise in field data.  Moreover, the new method preserves total energy, magnetic moment and phase space volume. \red{In its present version the method is designed for (quasi-)static electromagnetic fields.}

The integration procedure is based on a special 3D discretization of space resulting in \blue{piecewise}\red{locally} linear guiding-center
equations while retaining the symplectic property of the original set.
%\redx{with respect to piecewise linear fields}.
\red{
Formally such an interpolation could be denoted by Whitney forms or finite element exterior calculus in a way similar
to existing work on charged particle orbits~\cite{squire_geometric_2012,xiao_explicit_2015,kraus_gempic_2017} and results in a divergence-free magnetic field.
}
Within this discretization, vectors and
scalars characterizing the electromagnetic field are represented by continuous piecewise linear functions
which reduces the cost of spatial interpolation as compared to high-order interpolation with continuous derivatives
(e.g. with help of 3D-splines) required for usual integration. In return, the integration procedure requires
accurate tracing of intersections of the orbit with spatial cell boundaries where the coefficients of the linear
guiding-center equation set
are discontinuous. However, tracing of the boundaries is also required in Monte Carlo procedures for
the computation of the spatial distribution of various velocity space moments of the distribution function which are
computed as path integrals over the test particle dwell time within spatial cells.

In the original implementation of Ref.~\onlinecite{eder_three-dimensional_2019}, the linear guiding-center equation
set was solved numerically by using a usual Runge-Kutta (RK) integrator.
In this case, tracing of orbit intersections with spatial cell boundaries requires a few Newton iterations for the computation
of the dwell time within the cell. Since the integration error of this set scales with the third power of
the Larmor radius, accurate results can still be obtained by a single RK integration step over the dwell time.
For the same reason, sufficient accuracy can be achieved also by using the polynomial expansion of the solution over \red{an orbit parameter which allows to compute the dwell time} and integrals of velocity powers analytically.

In \red{the }case of magnetic fields with spatial symmetry (e.g., tokamaks with toroidal symmetry) the \red{quasi-}geometric integration
accurately preserves the respective (toroidal) canonical momentum. Thus, the property of such systems to ideally confine
the orbits is retained. In \red{the }absence of spatial symmetry, the parallel adiabatic invariant, which is an approximate conserved
quantity in stellarators, is well preserved, since the \red{quasi-}geometric integration does not lead to a significant error accumulation.

It should be noted that the piecewise linear approximation of the guiding-center equations represents field lines as
polygonal chains in coordinate space. Such a representation may introduce artificial chaos in case of 3D fields
when using non-aligned coordinate systems.
In our earlier publication~\citep{eder_three-dimensional_2019}, some field line diffusion has been observed
in a perturbed tokamak field which seemed to be in agreement with the quasilinear estimate assuming a continuous safety
factor profile of the unperturbed field (which is, actually, not the case), and, therefore, this diffusion has
been attributed to the linearization.
However, further detailed studies and resulting improvements of the algorithm revealed that the level of
artificial diffusion observed earlier is not connected with the intrinsic deficiency of the method. In the improved
algorithm this diffusion is actually much smaller so that it provides a negligible correction to the existing neoclassical
transport even for \blue{rather}\red{the relatively} strongly perturbed non-aligned 3D fields and a coarse mesh.

The above mentioned characteristics of the method allow, in particular, its effective application to the computation
of neoclassical transport coefficients using the Monte Carlo method. For demonstration, the mono-energetic neoclasscial
diffusion coefficient is evaluated here for a quasi-isodynamic reactor-scale stellarator
field~\cite{drevlak_quasi-isodynamic_2014}. The results and performance of the new method are compared to usual (RK)
orbit integration methods.

The remainder of this article is organized as follows. In section~\ref{sec:derivation_geometric_integrator} the spatial discretization procedure
is introduced and the numerical solution of the resulting piecewise linear guiding-center equations is described.
In section~\ref{sec:orbits} single particle orbits obtained with the \red{quasi-}geometric integration method and respective invariants of motion for axisymmetric and non-axisymmetric geometries are analyzed in detail. Furthermore, the introduced artificial chaos is studied.
The \blue{method's }application \red{of the method }to the evaluation of transport coefficients is presented in section~\ref{sec:transport}, where the computational orbit-integration performance is benchmarked as well.
Finally, the conclusion from the present study and further outlook is given in section~\ref{sec:summary}.
\red{Details on the Hamiltonian structure of the underlying locally linear equations are discussed in Appendix~\ref{sec:appendix_hamiltonian} and a useful feature for scoring orbit statistics is presented in Appendix~\ref{sec:appendix_integrals}.}

\section{Derivation of the orbit integration method and numerical solution}
\label{sec:derivation_geometric_integrator}

\subsection{Locally linear Hamiltonian guiding-center equations}
\label{sec:piecewise}

\blue{As a starting point, the equations of guiding-center motion\citep{boozer_guiding_1980} in general curvilinear coordinates $x^i$
are}
\red{As a starting point, the equations of guiding-center motion, Eq.~(11) of Ref.~\onlinecite{littlejohn_variational_1983} equivalent to Eq.~(A7) of Ref.~\onlinecite{boozer_guiding_1980}, are expressed in general curvilinear coordinates $x^i$,}

\be{eqm_curv}
\dot x^i = \frac{v_\parallel \varepsilon^{ijk}}{\sqrt{g}\, B_\parallel^\ast}\difp{A^\ast_k}{x^j},
\qquad A^\ast_k = A_k + \frac{v_\parallel}{\omega_c} B_k.
\ee

Here, $A_k$, $B_k$, $\omega_c$, $\Phi$ and $\sqrt{g}$ are the covariant components of the vector potential and the magnetic field, cyclotron frequency, electrostatic potential and the metric
determinant, respectively, and $\sqrt{g}\, B_\parallel^\ast = \varepsilon^{ijk} (B_i/B) \partial A^\ast_k/\partial x^j$.
Charge $e_\alpha$ and mass $m_\alpha$ of the considered species $\alpha$ enter
$\omega_c = e_\alpha B / (m_\alpha c)$ together with the magnetic field modulus
$B = \sqrt{B_k B^k}$ and the speed of light $c$.
The equations of motion are considered with the invariants $w=m_\alpha v^2/2+e_\alpha\Phi$ and $J_\perp=m_\alpha v_\perp^2/(2\omega_c)$ being total energy and perpendicular adiabatic invariant, respectively, and used as independent phase space variables. \red{The latter variable is related to the magnetic moment $\mu$ by a constant factor, $J_\perp=\mu m_\alpha c / e_\alpha$.}
The parallel velocity $v_\parallel$ in~\eq{eqm_curv} is not an independent variable but a known function of coordinates,
\be{defU}
v_\parallel^2=2U, \qquad
U=U(\bx)=\frac{1}{m}\left(w-J_\perp\omega_c(\bx)-e_\alpha\Phi(\bx)\right).
\ee
\redx{Due to the fact that the actual guiding-center orbits do not depend on the choice of their representing variables in phase space,
the parallel velocity $v_\parallel$ can also be treated as an independent variable. (Choosing $v_\parallel = v_\parallel \left(\bx, w, J_\perp \right)$ or $w = w \left(\bx, v_\parallel, J_\perp \right)$ is equivalent as long as Eq.~\eq{defU} is kept exact.)}
\bluex{Treating $v_\parallel$ as an independent variable now, i.e. replacing the first expression of~\eq{defU} by the differential equation}
\redx{By replacing the first expression of~\eq{defU} with the differential equation}
\be{}
\dot v_\parallel = \frac{1}{v_\parallel}\dot x^i \frac{\partial U} {\partial x^i},
\ee
the set (\ref{eqm_curv}) turns into
\bea{setofgceqs}
B_\parallel^\ast\sqrt{g}\,\dot x^i = \frac{\rd x^i}{\rd \tau}
&=&
\varepsilon^{ijk}\left(
v_\parallel \difp{A_k}{x^j}+2U\difp{}{x^j}\frac{B_k}{\omega_c}+\frac{B_k}{\omega_c}\difp{U}{x^j}
\right),
\nonumber\\
B_\parallel^\ast\sqrt{g}\,\dot v_\parallel = \frac{\rd v_\parallel}{\rd \tau}
&=&
\varepsilon^{ijk}\difp{U}{x^i}
\left(
\difp{A_k}{x^j}+v_\parallel \difp{}{x^j}\frac{B_k}{\omega_c}
\right).
\eea
\redx{Here, $v_\parallel$ should be treated as an independent variable only at its explicit occurrences,
while the quantity $U(\bx)$ should only be treated as a function of coordinates as defined by
the second expression of~\eq{defU}.}
Note that the invariants of motion remain in~\eq{setofgceqs} as parameters entering
the function of coordinates $U$\redx{.} \bluex{defined by the second expression of~\eq{defU}.}
\redx{As long as the contra-variant components of the phase space velocity defined by right hand sides
of set~\eq{setofgceqs} are used for obtaining the exact solution of this set, both
representations~\eq{defU} lead to the same result. However, if the derivatives of these components
are computed for the Jacobian as, e.g., in Appendix~\ref{sec:appendix_hamiltonian_Liouv}
the quantity $U$ should be treated as a function of parallel velocity defined by the first of~\eq{defU}.}
In~\eq{setofgceqs} the time variable is replaced
by an orbit parameter $\tau$ related to time by $\rd t = B_\parallel^\ast\sqrt{g}\rd \tau$,
and the time evolution is obtained implicitly from the integral $t(\tau)$.

\begin{figure}[h]
	\centerline{\includegraphics[keepaspectratio,width=1.2\linewidth, trim=50 100 50 0, clip]{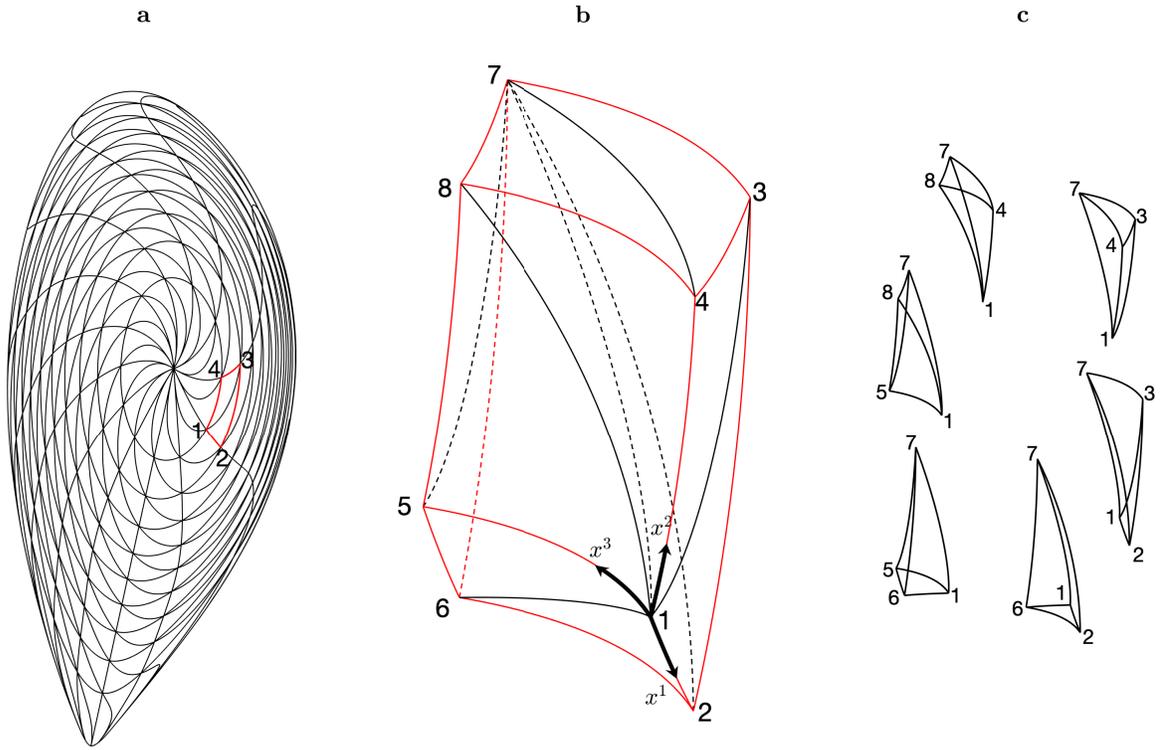}}
	\captionsetup{justification=raggedright,singlelinecheck=false,textfont=footnotesize,labelfont=footnotesize}
	\caption{Real space illustration of field-aligned grid for the plasma core of a toroidal fusion device. Magnetic field lines are traced in general curvilinear coordinates $(x^1$,$x^2$,$x^3)$ on $N_s$ flux surfaces. At equidistant spacing in toroidal and poloidal direction a 3D grid consisting of $N_s \times N_\vartheta \times N_\varphi$ hexahedra is created.\\
	(a) 2D poloidal projection of 3D grid with a marked hexahedron cell. (b) Magnification of marked hexahedron in (a) with indication how each hexahedron is split into six tetrahedral cells. \redx{The symmetry direction is along the coordinate $x^3$.}\blue{Resulting}\bluex{The resulting tetrahedra are }
%\redx{specially }
\blue{oriented in symmetry direction $x^3$ such that linear interpolation of electromagnetic field enables exact field representation on the cell's vertices.}
%\redx{does not introduce an interpolation error in symmetry direction $x^3$, in the case of axisymmetry. (Triangular tetrahedron face lying on $x^3=\mathrm{const.}$ is congruent with in the symmetry direction opposing face.)}
(c) Six individual tetrahedral cells compose stackable hexahedron such that adjacent tetrahedra faces are congruent.}
	\label{fig:curvilinear_grid}
\end{figure}

The special form~\eq{setofgceqs} allows to reduce computational effort and noise sensitivity by independently approximating
the field quantities $A_k$, $B_k/\omega_c$, $\omega_c$ and $\Phi$ by continuous piecewise linear functions.
Thus, curvilinear coordinate
space is split into tetrahedral cells with exact field values on the cell's vertices.
Fig. \ref{fig:curvilinear_grid} depicts such a \redx{real space illustration of a }curvilinear field-aligned grid for the plasma core of a tokamak. \bluex{, where splitting along the symmetry direction is achieved with stackable hexahedra, each consisting of six tetrahedral cells.}
\redx{These tetrahedral cells must be specially oriented (explained below in section~\ref{sec:orbits_2d}) in order to preserve an invariant of motion in the case of axisymmetry.}

As a result \redx{of this piecewise field linearization}, in each cell, the equations of motion (\ref{setofgceqs})
turn into a set of four linear ODEs with constant coefficients
\be{standeqset}
\frac{\rd z^i}{\rd \tau} = a^i_l z^l + b^i,
\ee
in phase-space variables $z^i=x^i$ for $i=1,2,3$ and $z^4=v_\parallel$. The matrix elements are
\bea{amatdef}
a^i_l &=& \varepsilon^{ijk}\left(
2\difp{U}{x^l}\difp{}{x^j}\frac{B_k}{\omega_c}+\difp{U}{x^j}\difp{}{x^l}\frac{B_k}{\omega_c}
\right)
\qquad\mbox{for}\qquad 1\le i,l \le 3,
\nonumber \\
a^i_4 &=& \varepsilon^{ijk} \difp{A_k}{x^j}
\qquad\mbox{for}\qquad 1\le i \le 3,
\nonumber \\
a^4_l &=& 0
\qquad\mbox{for}\qquad 1\le l \le 3,
\nonumber \\
a^4_4 &=& \varepsilon^{ijk}\difp{U}{x^i}
\difp{}{x^j}\frac{B_k}{\omega_c},
\eea
and components of vector $b^i$ are
\bea{bvecdef}
b^i &=& \varepsilon^{ijk}\left(
2U_0\difp{}{x^j}\frac{B_k}{\omega_c}+\left(\frac{B_k}{\omega_c}\right)_0\difp{U}{x^j}
\right)
\qquad\mbox{for}\qquad 1\le i \le 3,
\nonumber \\
b^4 &=& \varepsilon^{ijk}\difp{U}{x^i}
\difp{A_k}{x^j},
\eea
where quantities with zero mean the value at the origin of the coordinates \red{shifted in each
tetrahedral cell to one of the cell's vertices,}

\be{valorig}
U=U_0+x^i\difp{U}{x^i}, \qquad \frac{B_k}{\omega_c} = \left(\frac{B_k}{\omega_c}\right)_0
+x^i \difp{}{x^i}\frac{B_k}{\omega_c}.
\ee

%\red{In each tetrahedral cell a local coordinate system is used with the coordinate origin lying in one of the cell's vertices. %Thus, terms including the coordinate origin $x^i_0$ vanish and are not explicitly written in the above representation of the %ODEs' coefficients.}

Since the piecewise constant coefficients of set~\eq{standeqset} are discontinuous at
the cell boundaries, orbit intersections with tetrahedra faces
must be computed exactly when integrating particle trajectories.

In fact, a linear approximation of field quantities which locally breaks the physical connection between them
does not destroy the Hamiltonian nature of the original set~\eq{eqm_curv}. Indeed, despite the
approximation made, equation set~\eq{setofgceqs} can still be cast to the non-canonical Hamiltonian
form
\be{hamform}
\frac{\rd z^i}{\rd \tau}=\Lambda^{ij}\difp{H}{z^j}, \qquad \Lambda^{ij}(\bz)=\left\{z^i,z^j\right\}_\tau,
\ee
where the Hamiltonian function is $H(\bz)=v_\parallel^2/2-U(\bx)$ and $\Lambda^{ij}(\bz)$ is an antisymmetric Poisson matrix.
The latter is linked to Poisson brackets that are slightly re-defined from those in
Ref.~\onlinecite{grebogi_relativistic_1984} due to a different orbit parameter,
\be{poisson}
\left\{f,g\right\}_\tau = b_\ast^i\left(\difp{f}{x^i}\difp{g}{v_\parallel}-\difp{g}{x^i}\difp{f}{v_\parallel}\right)
+\varepsilon^{ijk}\difp{g}{x^i}\difp{f}{x^j}\frac{B_k}{\omega_c},
\qquad
b_\ast^i = \varepsilon^{ijk}\left(\difp{A_k}{x^j}+v_\parallel \difp{}{x^j}\frac{B_k}{\omega_c}\right).
\ee
In the derivation above, one occurrence of $v_\parallel^2$ has been replaced by $2U(\bx)$ to obtain equation set~\eq{setofgceqs}.
As long as this equality in the first of~\eq{defU} is kept exact by a numerical scheme,
\redx{this formal violation of the Poisson structure doesn't affect the final result.
The present method integrates equations of motion to computer accuracy and thus
exactly conserves invariants with respect to piecewise linear fields in order
to fulfil this requirement. Invariants with respect to original smooth fields
oscillate within fixed bounds in a similar way as for conventional symplectic
integrators~\cite{hairer_geometric_2006}.}
\bluex{a geometric Poisson integrator follows,
being a generalization of symplectic integrators in non-canonical coordinates. Similar to symplectic integrators, Poisson integrators
preserve the symplectic structure and invariants of motion within fixed bounds.~\cite{hairer_geometric_2006}}
\red{
Conservation properties and symplectic features of the locally linear
equation set \eq{standeqset} are discussed in more detail in Appendix~\ref{sec:appendix_hamiltonian}.
}

\subsection{Numerical solution}
\label{sec:numerical_solution}

An approximate formal solution of set~\eq{standeqset} in a single cell is given as a polynomial series of the orbit parameter,
\be{explser}
\bz(\tau)=\bz_0 + \sum\limits_{k=1}^K \frac{\tau^k}{k!} \left(\hat \ba^{k-1}\cdot\bb + \hat \ba^k\cdot \bz_0\right),
\ee
where $\hat \ba$ and $\bb$ stand for matrix $a^i_l$ and vector $b^i$, respectively,
$\bz(0)=\bz_0$ is a starting point,
and the exact solution is obtained in the limit $K \rightarrow \infty$. \red{In the case $k=1$ the matrix $\hat \ba^{k-1} = \hat \ba^{0}$ is the identity matrix.}
\redx{It should be noted that for mild electric fields (validity domain of Eq.~\eq{eqm_curv})
where the potential energy $e_\alpha\Phi$ is of the order of the kinetic energy (sub-sonic rotations)
all elements of matrix $\hat\ba$ except for $a^i_4$ scale linearly with the Larmor radius, and
in the zero Larmor radius limit the series expansion of the order $K =  2$ does already provide an exact solution.
Therefore, the series expansion~\eq{explser} converges rapidly in the case that the Larmor radius is small in comparison with
 the spatial scale of the electromagnetic field (see Eq.~\eq{RK4estimation} below).}

%\red{It should be mentioned that particle positions, including the starting point $\bz_0$, are not limited to cell boundaries but %can exist continuously in the relevant coordinate space. This implies that for a given particle position also the appropriate cell %which encloses this position, and thus defines the coefficients of Eq.~\eq{explser}, must be identified.}

Since intersections with cell boundaries (tetrahedra faces) must be computed exactly,
the particle is pushed from cell boundary to cell boundary, if no intermediate position inside a cell,
e.g. after a pre-defined time step, is required deliberately.
\red{In the latter case, the positions before and after the time step are located inside the cells, while the time
step itself includes tracing of all orbit intersections with cell boundaries between these positions.}

\begin{SCfigure}
	\centering{	\includegraphics[width=0.5\textwidth,keepaspectratio,trim=200 100 200 100, clip]{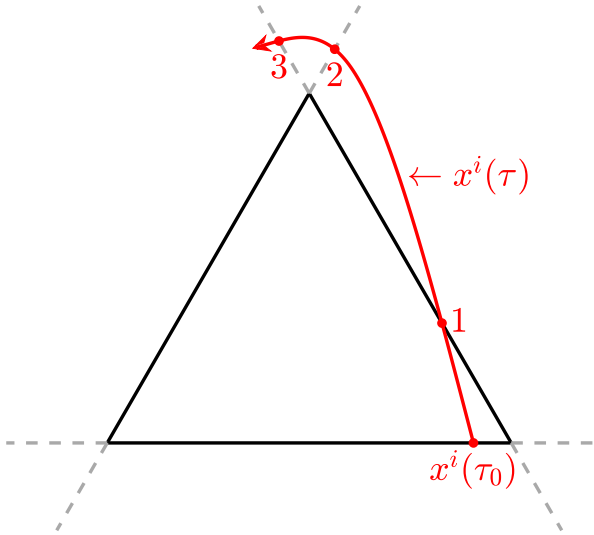}}
	\captionsetup{textfont=footnotesize,labelfont=footnotesize,justification = raggedright,singlelinecheck = false}
	 \caption{Illustration of intersections of the orbit $x^i(\tau)$ with planes confining the cell. \blue{For simplification, the }\red{The} three-dimensional tetrahedral cell is depicted as a two-dimensional triangle\blue{.}\red{ in the interest of simplification.} The particle enters the cell at $x^i(\tau_0)$ and leaves the cell at \raisebox{.5pt}{\textcircled{\raisebox{-.9pt} {1}}}. Other intersections at the points \raisebox{.5pt}{\textcircled{\raisebox{-.9pt} {2}}} and \raisebox{.5pt}{\textcircled{\raisebox{-.9pt} {3}}} are not realized. The correct orbit-tetrahedron intersection~\raisebox{.5pt}{\textcircled{\raisebox{-.9pt} {1}}} is reached in the smallest positive time among all intersections.}
	\label{fig:triangle01}
\end{SCfigure}

An orbit intersection with a tetrahedron is found on exit as an intersection with one of
four planes
\be{planes}
F^\alpha(\bz)\equiv n^{(\alpha)}_i \left(x^i-x^i_{(\alpha)}\right)=0, \qquad \alpha=1,\dots,4
\ee
reached in the smallest positive time,\blue{after entry (``dwell time'')} \red{``exit time'', from the starting position $x^i(\tau_0)$ located either on the cell boundary or inside the cell }, see Fig.~\ref{fig:triangle01}.
Here, $n^{(\alpha)}_i$ and $x^i_{(\alpha)}$ are the (constant)
normal to the plane containing tetrahedron face $\alpha$ and coordinates of some vertex
on that face, respectively.
\red{In the case that the starting position is located on the cell boundary (due to boundary-boundary-pushings as in Fig.~\ref{fig:triangle01}), the exit time coincides with the ``dwell time'' of the particle inside the cell. If an intermediate stop (inside the cell) is deliberately required, the dwell time is the sum of the time to reach this stop (``entry time'') and the exit time.}

With substitution of the orbit, $\bz=\bz(\tau)$, Eqs.~\eq{planes}
are nonlinear equations with respect to the orbit parameter, $F^\alpha(\bz(\tau))=0$,
which should be solved numerically. They become algebraic and can be solved analytically
if an approximate solution~\eq{explser} is used with $K\le 4$\red{, as explained below in detail.}

Both cases have been implemented in a Fortran program with the name \blue{Geometric}\red{\textbf{G}uiding-center}
\textbf{OR}bit \textbf{I}ntegration with \textbf{L}ocal \textbf{L}inearization \textbf{A}pproach
(\bluex{\textit{Gorilla}}\redx{\textbf{GORILLA}}).

\redx{As a matter of completeness, it is worth mentioning that a third similar to Ref.~\onlinecite{kasilov_geometric_2016} variant exists
to solve the linear equation set~\eq{standeqset} exactly in terms of exponential functions
of eigenvalues in the eigenvector basis (as any other linear equation set). Formally, this solution
corresponds to the limit $K \rightarrow \infty$ in Eq.~\eq{explser} where infinite sums
can be expressed in terms of the exponential function $\exp(\tau \hat a)$. This method, however,
has two drawbacks. First, Eqs.~\eq{planes} turn into nonlinear transcendental equations
which have to be solved numerically up to computer accuracy. Second, the analytical solution results
in strong cancellation errors in the case of small Larmor radii which is exactly the case where the finite
series expansion results in negligible errors. Therefore, we refrain from such a solution.}

In its first variant (\bluex{\textit{Gorilla RK4}}\redx{GORILLA RK4}), the orbit intersections with tetrahedra faces are
computed numerically by solving equation set (\ref{standeqset}) in each cell with a single step of
the Runge-Kutta 4 method embedded into an iterative scheme
using Newton's
method to obtain the integration time $\Delta\tau$ required to reach the cell boundary
(\red{exit time $\tau_e=\Delta \tau$}\blue{dwell time $\tau_d=\Delta \tau$}).
\blue{Predominantly, this}\red{This} iterative scheme \red{predominantly }converges after roughly two Newton iterations, due to an
analytic estimation for the necessary initial
step length using an approximate parabolic solution of ODE set (\ref{standeqset}) taken in zeroth order
in Larmor radius, $a^i_l = 0$ for $1 \le i,j \le 3$ and $a^4_4 v_\parallel(\tau) = a^4_4 v_{\parallel,0}$.

Nevertheless, in \red{the }case of numerically challenging orbits (tangential to a tetrahedron face or
almost intersecting with a tetrahedron's edges or vertices),
several special cases appear \blue{where}\red{in which} the iterative scheme \blue{doesn't}\red{does not} converge, and those cases must be treated
separately in a computationally more expensive manner.
Since such cases appear only rarely, the \blue{additional}\red{additionally} required computational effort is negligible in
comparison to the standard \blue{one}\red{procedure}.

A single RK4 integration step per iteration is sufficient because the magnetic field is uniform
within a cell. Respectively, the error of the RK4 method strongly
scales with the Larmor radius $\rho$ and can be brought below computer accuracy by a moderate
grid refinement.
Namely, in a tokamak geometry the error of a single step traversing the cell can be estimated as
\be{RK4estimation}
\frac{\delta R(\Delta\tau)}{a} \sim \frac{\rho^3}{q^4 R^3} \Delta \varphi^5,
\ee
with $R$, $a$, $q$ and $\Delta\varphi$ denoting major radius, plasma radius, safety
factor and toroidal cell length, respectively.

Due to the scaling of the error with $\rho$, for particles with mild energies (thermal electrons
and ions) a less elaborate algorithm (\bluex{\textit{Gorilla Poly}}\redx{GORILLA Poly}) can be realized.
By truncating the summation of Eq.~\eq{explser} at $K$~=~2, 3~or~4 one obtains
approximate solutions of various orders in Larmor radius.
With these solutions, equations for the \blue{dwell time $\tau_d$}\red{exit time $\tau_e$}
are algebraic equations which are solved analytically up to order $K=4$, by finding the
smallest positive root of
\be{root_explser}
\bn^\alpha \cdot \left( \bz_0 + \sum\limits_{k=1}^K
\frac{\tau_{\red{e}\blue{d}}^k}{k!} \left(\hat \ba^{k-1}\cdot\bb + \hat \ba^k\cdot \bz_0\right)
-\bz^\alpha
\right) = 0,
\ee
where $\bn^\alpha = \left(n_1^{(\alpha)},n_2^{(\alpha)},n_3^{(\alpha)},0\right)$
are face normals (see above) and
$\bz^\alpha=\left(x^1_{(\alpha)},x^2_{(\alpha)},x^3_{(\alpha)},0\right)$.
Furthermore, by using the quadratic polynomial solution $K$~=~2 of Eq.~\eq{root_explser},
the appropriate orbit intersection plane $\alpha$ can be predicted for higher orders
$K$ = 3 or 4, which predominantly reduces the number of higher order root finding operations
from four to one.
In the numerical implementation,
series~\eq{explser} which contains matrix products is not directly evaluated as written above. Instead, various sub-products
are preliminarily evaluated and stored for grid cells\red{,} \blue{which minimizes}\red{minimizing} the number of matrix products
\blue{needed}\red{that need} to be evaluated directly.
The finite series solution~\eq{explser}
allows to analytically evaluate also various integrals
over the dwell time needed for scoring of macroscopic plasma plasma parameters in Monte Carlo
procedures.
In Appendix~\ref{sec:appendix_integrals}, such integrals of $v_\parallel$ and
of $v_\perp^2$ and $v_\parallel^2$ are given,
which respectively determine parallel plasma flow and
components of the pressure tensor in Chew-Goldberger-Low form
essential for computation of equilibrium plasma currents.

\section{Collisionless guiding-center orbits}
\label{sec:orbits}

\subsection{Guiding-center orbits in an axisymmetric tokamak field}
\label{sec:orbits_2d}

In this section the results of \red{quasi-}geometric orbit integration computed with \bluex{\textit{Gorilla}}\redx{GORILLA} and the comparison to an exact orbit computed with a usual adaptive RK4/5 integrator are presented for an axisymmetric tokamak field configuration of ASDEX Upgrade (shot 26884 at 4300 ms) described in Ref.~\onlinecite{heyn_quasilinear_2014}. The adaptive RK4/5 integrator requires high-order interpolation of electromagnetic fields with continuous derivatives, e.g. with help of 3D-splines, instead of continuous piecewise linear functions \red{as }in \red{the }case of the \red{quasi-}geometric integration method.

In axisymmetric configurations, the shape of the orbit is fully determined by three conservation laws,
$p_\varphi = \mathrm{const.}$, $J_\perp = \mathrm{const.}$ and $w =  \mathrm{const.}$, where
\be{ptor}
p_\varphi\redx{ = \frac{e}{c} A_\varphi^\ast} = m v_\parallel \frac{B_\varphi}{B} + \frac{e}{c}A_\varphi
\ee
is \redx{the} canonical toroidal angular momentum. \redx{The conservation of $p_\varphi$ is obvious from Eq.~\eq{eqm_curv}
since after its substitution in $\dot p_\varphi=\dot x^i \partial p_\varphi / \partial x^i$ all the
terms in the resulting expression are proportional to partial derivatives over $\varphi$ of various $A^\ast_k$ components.}

In geometric/symplectic numerical integration schemes \bluex{this property is}\redx{these conservation properties are} retained [\onlinecite{hairer_geometric_2006}], which means that orbits must remain closed in the poloidal projection.

\bluex{Due to the special formulation of Eq.~\eq{defU}, where $v_\parallel$ is purely a function of position, the perpendicular adiabatic invariant and the total energy are conserved naturally.} \bluex{By properly orienting the tetrahedra with respect to the symmetry direction, axisymmetry in \red{the }case of 2D fields
is exactly preserved upon linearization.}
\redx{
In order to preserve $p_\varphi$ upon linearization, the tetrahedral cells of the method's underlying
grid must be specially oriented with respect to the symmetry direction. Namely, this is achieved with
stackable hexahedra each consisting of two triangular prisms which are both subsequently split into
3 tetrahedral cells.  The prisms are oriented such that all triangular prism faces lie on
$x^3= \mathrm{const.}$ planes where $x^3$ is the symmetry direction.
Thus, each tetrahedron face which lies on a $x^3= \mathrm{const.}$ plane is congruent with
all other tetrahedron faces opposing it in the symmetry direction.
This specific splitting realization can be seen in Fig.~\ref{fig:curvilinear_grid}.
}
Consequently, the canonical toroidal angular momentum remains invariant \redx{in the presented method}\redx{, since the linearization of the electromagnetic field within a tetrahedral cell does not introduce an interpolation error in the symmetry direction. Namely, partial derivatives of the field quantities with respect to the symmetry direction remain zero. In the following, cylindrical ($R$, $\varphi$, $Z$)
and symmetry flux coordinates ($s$, $\vartheta$, $\varphi$) of Ref.~\onlinecite{dhaeseleer_flux_2012} are used, where $\varphi = x^3$ in both coordinate systems.}

\begin{figure}[h]
	\centerline{\includegraphics[keepaspectratio,width=1.2\linewidth]{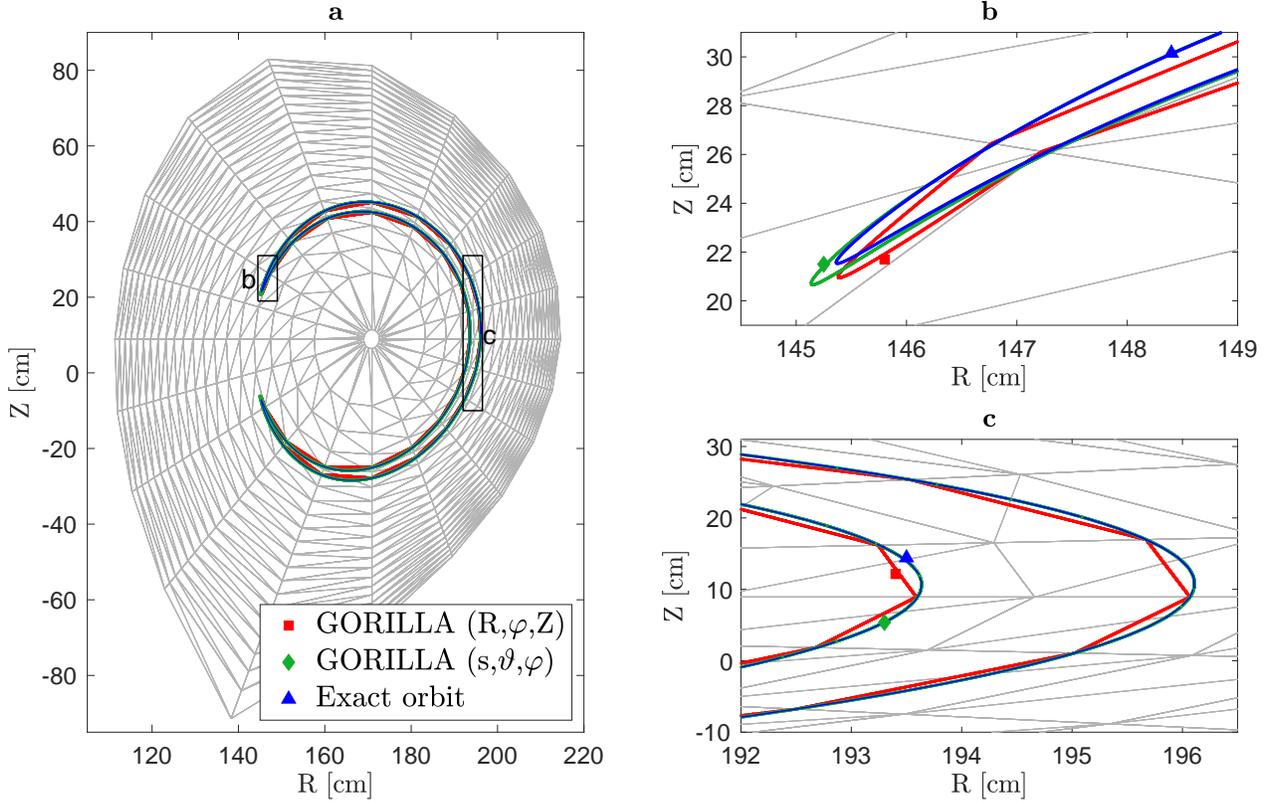}}
	\captionsetup{justification=raggedright,singlelinecheck=false,textfont=footnotesize,labelfont=footnotesize}
	\caption{(a) Poincaré plot ($\varphi=0$, $10^7$ toroidal mappings) of a trapped 1.5~keV D ion in axisymmetric ASDEX Upgrade configuration with a tetrahedral grid size of 20x20x20. Two-dimensional Poincaré sections of orbits obtained with different integration methods are indicated with markers: Exact \blue{orbit:}\red{orbit (adaptive RK4/5):} $\blacktriangle$, \blue{3D Geometric Integrator }\bluex{\textit{Gorilla}}\redx{GORILLA} with cylindrical coordinates: $\blacksquare$, \bluex{\textit{Gorilla}}\redx{GORILLA} with symmetry flux coordinates: $\blacklozenge$. (b) and (c) are magnifications of the pertinent zones in (a).}
	\label{fig:orbit_comparison}
\end{figure}

Fig. \ref{fig:orbit_comparison} depicts Poincaré plots ($\varphi = 0$) of trapped thermal ion orbits making $10^7$ toroidal turns which are integrated
by different methods from the same starting conditions. \blue{G}\red{Quasi-g}eometric integration using an iterative scheme with RK4 integration and Newton steps has been performed in cylindrical\bluex{($R$, $\varphi$, $Z$)} and symmetry flux coordinates\bluex{ ($s$, $\vartheta$, $\varphi$) of Ref.~\onlinecite{dhaeseleer_flux_2012}} and is compared to the exact orbit.
It can be seen, that the coarseness of the grid leads to slightly differently shaped orbits obtained in different coordinate systems, whereas the
effect of the integration error (\ref{RK4estimation}) is negligible.

\begin{figure}[h]
	\centerline{\includegraphics[keepaspectratio,width=1.1\linewidth,trim=60 0 60 0, clip]{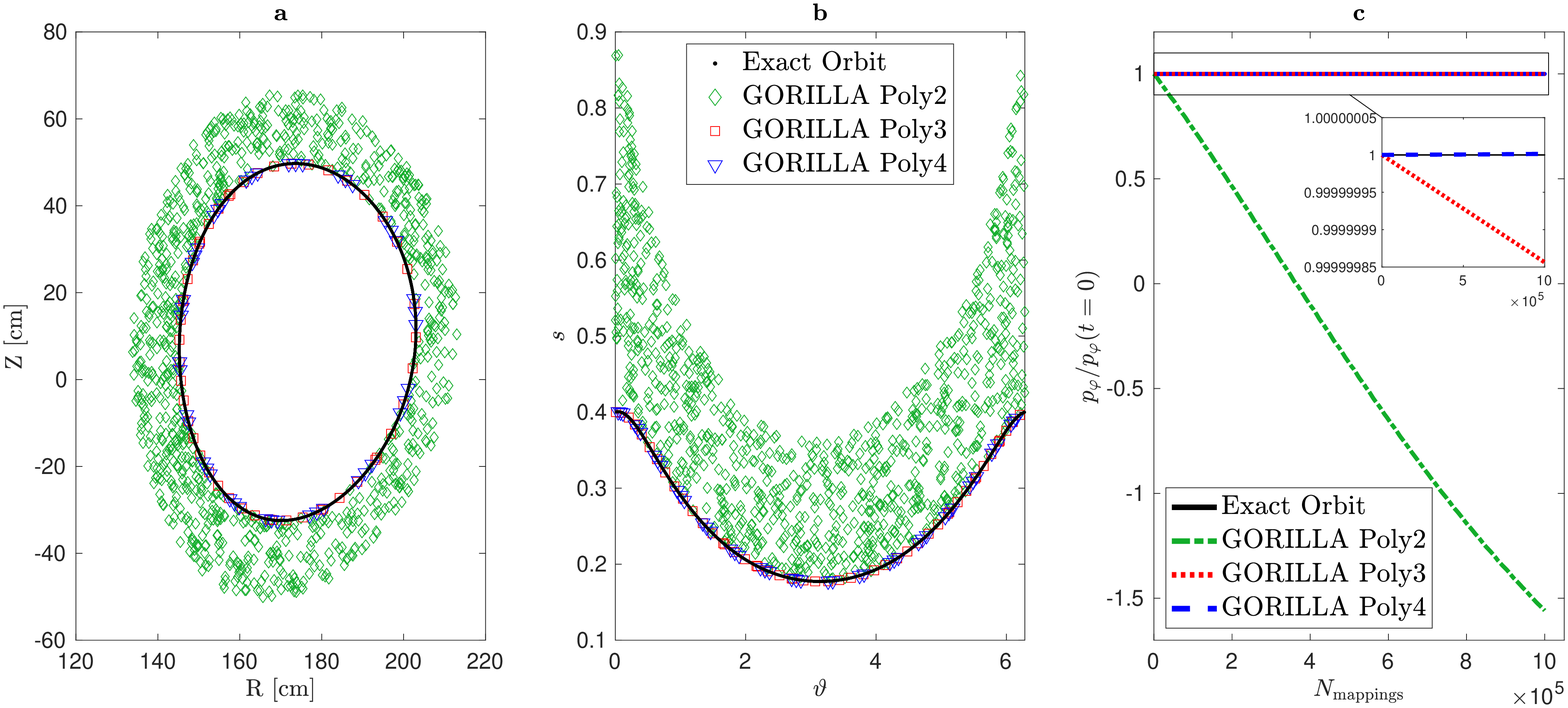}}
	\captionsetup{justification=raggedright,singlelinecheck=false,textfont=footnotesize,labelfont=footnotesize}
	\caption{(a \& b) Poincaré plot ($\varphi=0$, $10^6$ toroidal mappings) of a passing 300 keV D ion in axisymmetric ASDEX Upgrade configuration evaluated by \blue{the 3D Geometric Integrator }\bluex{\textit{Gorilla}}\redx{GORILLA} with the analytical solution in form of a polynomial series truncated at $K=2$ ($\Diamond$), $K=3$ ($\square$) and $K=4$ ($\triangledown$) compared to the exact orbit ($\bullet$). Cylindrical coordinates are used for the Poincaré plot in (a), whereas symmetry flux coordinates are used in (b).\\
		(c) Canonical toroidal angular momentum $p_\varphi$ normalized to the value at $t=0$ is evaluated at the Poincaré sections in (a): Exact result (solid) is compared to polynomial series truncated at $K=2$ (dash-dotted), $K=3$ (dotted) and $K=4$ (dashed).}
	\label{fig:tokamak_passing_toroidal_angular}
\end{figure}

In Fig.~\ref{fig:tokamak_passing_toroidal_angular} results of the \red{quasi-}geometric integration of a passing high energy ion (300~keV) using the polynomial series solution~\eq{explser} are shown for the same axisymmetric geometry and compared to the exact orbit.
Symmetry flux coordinates are used for the \red{quasi-}geometric integration, the results are then converted to cylindrical coordinates
in the first plot (a). In general, the quartic polynomial solution ($K$~=~4) of ODE set~\eq{standeqset} is equivalent to the
numerical solution using Runge-Kutta 4, thus, the result of the latter is omitted in the figure.
Moreover, it can be seen that even for high
energy ions the series expansions of third and fourth order are already accurate enough in order to fulfill the condition
$p_\varphi=\text{const.}$ over 10$^6$~toroidal mappings. However, the second order series expansion shows a convective behavior, due to a systematic error from solving the ODE set~\eq{standeqset}.
\blue{Nevertheless, for the electrons which have much smaller Larmor radii, the second order series expansion is sufficient.}\red{Nevertheless, the second order series expansion is sufficient for electrons which have much smaller Larmor radii.}

\begin{figure}[t]
	\centerline{\includegraphics[keepaspectratio,width=1.2\linewidth]{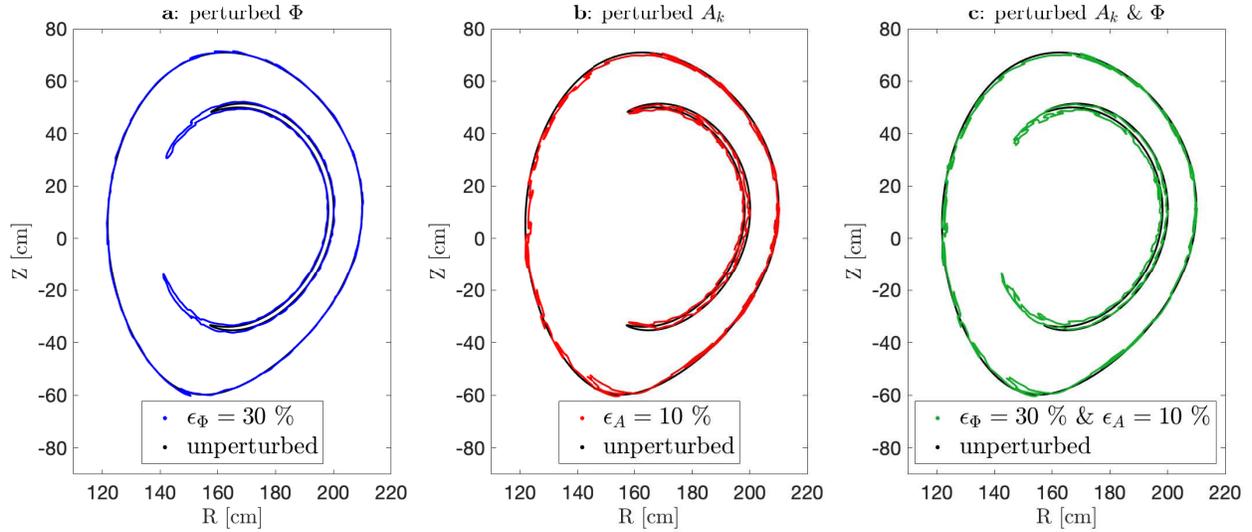}}
	\captionsetup{justification=raggedright,singlelinecheck=false,textfont=footnotesize,labelfont=footnotesize}
	\caption{Poincaré plot ($\varphi=0$) of trapped and passing 1.5 keV D ions in axisymmetric ASDEX Upgrade configuration with axisymmetric perturbation of electrostatic potential $\Phi$ and vector potential $A_k$. \red{Orbits are evaluated by \bluex{\textit{Gorilla}}\redx{GORILLA} with the analytical solution in form of a polynomial series truncated at $K=4$.}\\
The Poincaré plot for unperturbed electromagnetic fields is depicted in black as a reference. Uniformly distributed axisymmetric random noise ($\xi = 0 \dots 1$) is added in (a) to the electrostatic potential $\Phi^{\text{noisy}} = \Phi(1+\epsilon_\Phi \xi)$, in (b) to the vector potential	$A_k^{\text{noisy}} = A_k(1+\epsilon_A \xi)$ and in (c) to both quantities.
}
	\label{fig:axisymmetric_noise_plot}
\end{figure}

To obtain the orbits of thermal ions shown in Fig.~\ref{fig:axisymmetric_noise_plot}, uniformly distributed axisymmetric random noise ($\xi = 0 \dots 1$) is added to the electrostatic potential $\Phi^{\text{noisy}} = \Phi(1+\epsilon_\Phi \xi)$, to the vector potential $A_k^{\text{noisy}} = A_k(1+\epsilon_A \xi)$ and simultaneously to both quantities, respectively. Here, $\epsilon$ is the relative magnitude of added noise. \red{The guiding-center orbits are evaluated in symmetry flux coordinates by \bluex{\textit{Gorilla}}\redx{GORILLA} with the analytical solution in form of a polynomial series truncated at $K=4$, the results are then converted to cylindrical coordinates.}
Even though relatively high noise (up to 30~\%) is added, the orbits keep a similar shape in comparison with the unperturbed ones and remain closed in the poloidal projection, meaning the condition $p_\varphi = \text{const.}$ is still fulfilled.

Therefore, the \red{quasi-}geometric integration method is suitable for self-consistent Monte Carlo modeling of 2D equilibrium plasma
parameters and electromagnetic fields computed from those parameters.
Further, it should be noted that the computational efficiency of the \red{quasi-}geometric integrator is not affected
by the presence of small-scale noise. However, this would be the case for an adaptive RK4/5 integration in combination
with high order smooth interpolation where the noise leads to small-scale oscillations.

\begin{figure}[h]
	\centerline{\includegraphics[keepaspectratio,width=1.2\linewidth]{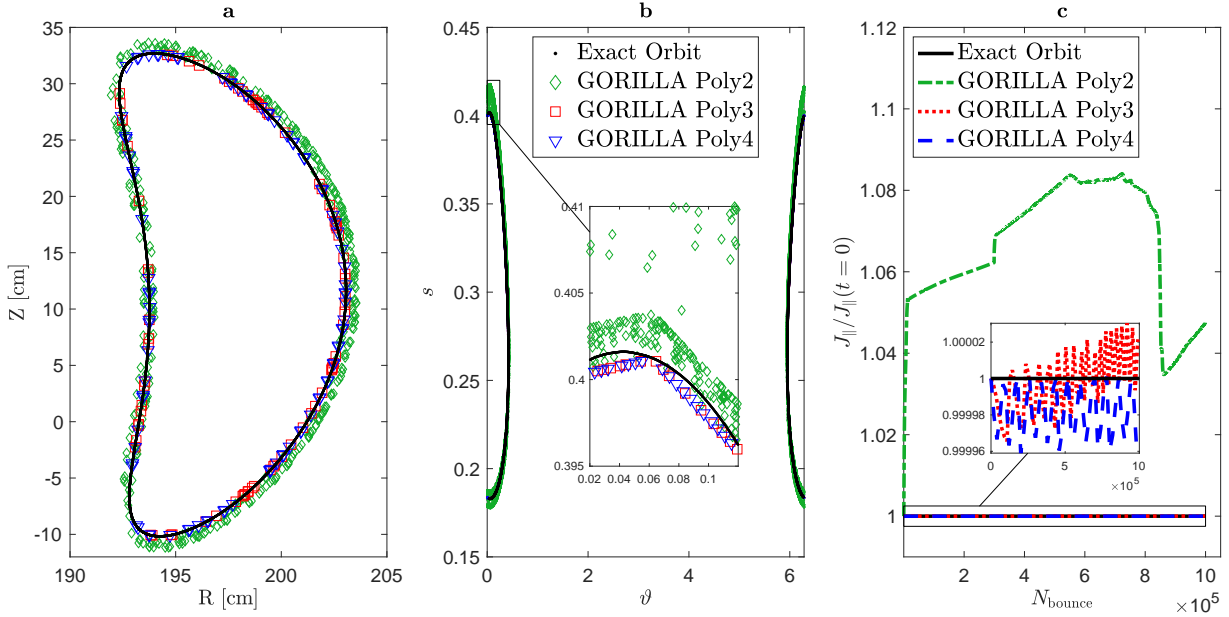}}
	\captionsetup{justification=raggedright,singlelinecheck=false,textfont=footnotesize,labelfont=footnotesize}
	\caption{(a \& b) Poincaré plot ($\varphi=0$, $10^6$ bounce periods) of a trapped 300 keV D ion in axisymmetric ASDEX Upgrade configuration evaluated by \blue{the 3D Geometric Integrator }\bluex{\textit{Gorilla}}\redx{GORILLA} with the analytical solution in form of a polynomial series truncated at $K=2$ ($\Diamond$), $K=3$ ($\square$) and $K=4$ ($\triangledown$) compared to the exact orbit ($\bullet$). Cylindrical coordinates are used for the Poincaré plot in (a), whereas symmetry flux coordinates are used in (b). The magnification in (b) shows the banana tip and clearly depicts the difference of polynomial orders. \\
	(c) The parallel adiabatic invariant $J_\parallel$ normalized to the value at $t=0$ is depicted for $10^6$ bounce periods: Exact result (solid) is compared to polynomial series truncated at $K=2$ (dash-dotted), $K=3$ (dotted) and $K=4$ (dashed).}
	\label{fig:tokamak_trapped_parallel_invariant}
\end{figure}

The Poincaré plots ($\varphi = 0$) of Fig.~\ref{fig:tokamak_trapped_parallel_invariant} (a) \& (b) correspond to a trapped high energy ion (300~keV), where the orbits are integrated using the polynomial series expansion in several orders and compared to the exact orbit. Again, symmetry flux coordinates are used for the \red{quasi-}geometric integration, the results are then converted to cylindrical coordinates in the first plot (a). Despite the same starting conditions \blue{of}\red{for} all orbits, the magnification in (b) clearly depicts differences in the shape of the orbits caused by a finite grid size ($100 \times 100 \times 100$) and different orders of the series expansion.

Fig.~\ref{fig:tokamak_trapped_parallel_invariant} (c) shows the corresponding time evolution of the parallel adiabatic invariant~$J_\parallel$, which is defined by an integral over the distance $l$ passed along the field line during a single bounce period $t_b$ by a
trapped particle as follows,
\be{parinv}
J_\parallel = m \oint v_\parallel \mathrm{d}l = m \int_{0}^{t_b} v_\parallel^2 \left(t \right) \mathrm{d}t.
\ee
For an exact orbit, $J_\parallel$ is a conserved quantity. Symplectic orbit integration does not lead to an error accumulation in the invariants of motion [\onlinecite{hairer_geometric_2006}], while systematic changes can only arise from numerical errors in solving the ODE. Even for high energy ions the series expansions of third and fourth order are already accurate enough in order to fulfill the condition $J_\parallel = \mathrm{const.}$ for $10^6$ bounce periods. Hence, in an axisymmetric configuration excellent long-term orbit dynamic is demonstrated by the \red{quasi-}geometric orbit integration method, as long as the ODE set~\eq{standeqset} is solved accurately.

\subsection{Guiding-center orbits in three-dimensional fields}
\label{sec:orbits_3d}

It can be seen from Eqs.~\eq{standeqset} and~\eq{amatdef} that in the field line limit $\omega_c\rightarrow\infty$
guiding-center orbits are straight within spatial cells and magnetic field lines are represented by polygonal chains, respectively.
In case of 3D magnetic fields described in non-aligned spatial coordinates, the existence of embedded KAM surfaces
is not obvious for such an approximate representation of the field even in \red{the }case \red{where }these surfaces exist in the exact
system. In order to study artificial chaos induced by the linearization,
\red{quasi-}geometric orbit integration has been performed for low energy particles with negligible FLR effects in symmetry flux coordinates
$(s,\vartheta,\varphi)$
associated with the axisymmetric tokamak field of the previous section with a harmonic perturbation added
to the toroidal co-variant component of the axisymmetric vector potential,
\be{example_onemode}
A_\varphi=\psi_{\rm pol}(s)(1+\varepsilon_M\cos(m_0\vartheta+n_0\varphi)).
\ee
The harmonic indices $m_0=n_0=2$ used in the testing correspond to a non-resonant perturbation
which leads only to a corrugation of the magnetic surfaces\red{,} \blue{such}\red{with the effect} that they are \blue{not aligned}\red{no longer aligned} with the coordinate
surfaces $s=\mathrm{const.}$\blue{ anymore}.

\begin{figure}[h]
	\centerline{\includegraphics[keepaspectratio,width=1.2\linewidth]{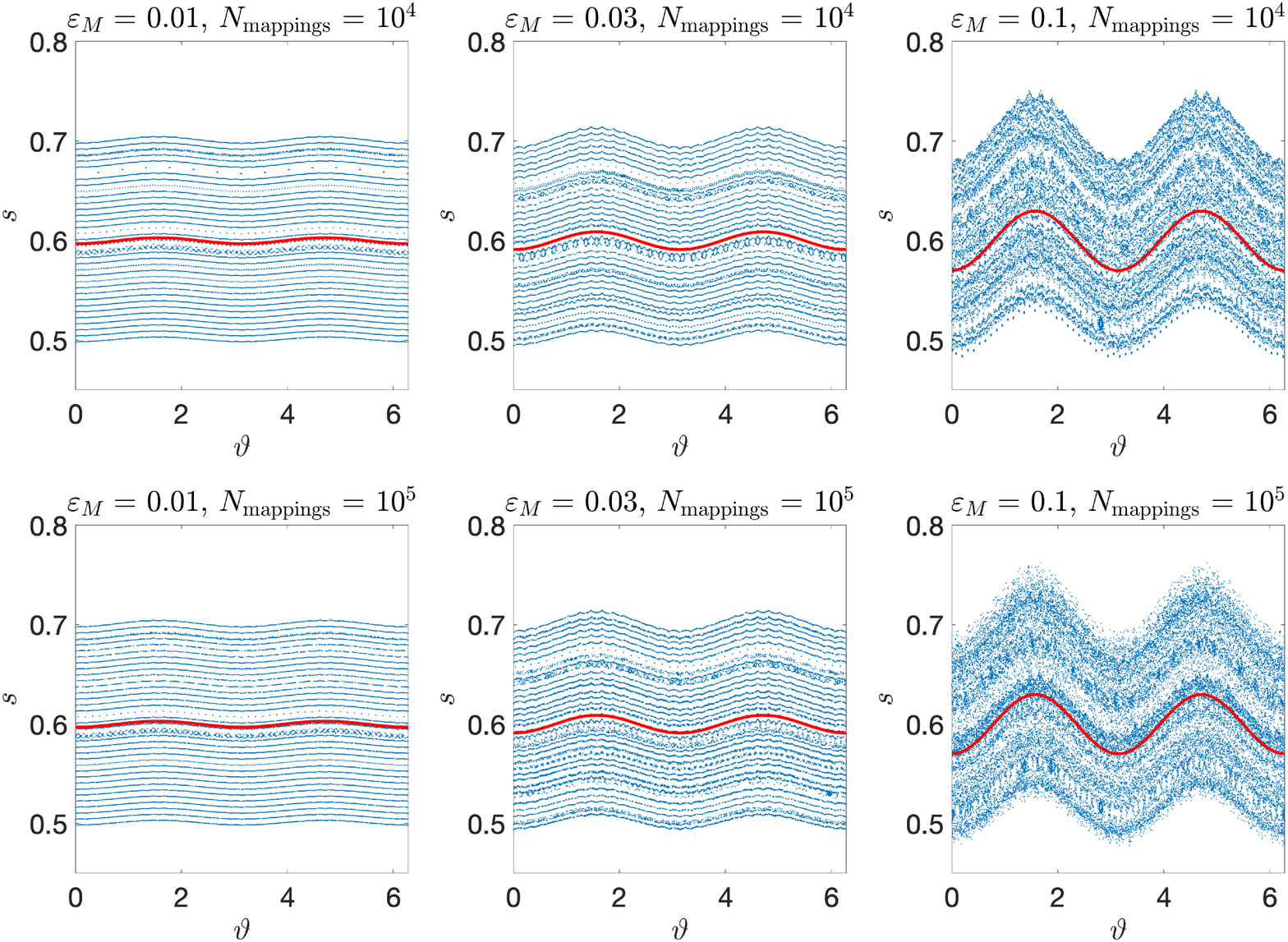}}
	\captionsetup{justification=raggedright,singlelinecheck=false,textfont=footnotesize,labelfont=footnotesize}
	\caption{Poincaré plots of the orbits in zero Larmor radius limit (field lines)
	for $10^4$ (upper row) and $10^5$ (lower row) toroidal mappings and various perturbation amplitudes indicated
	in the titles.
	Orbits start at 34 equidistant flux surfaces between $s = 0.5$ and  $s=0.7$ and are evaluated by \bluex{\textit{Gorilla}}\redx{GORILLA}
	with polynomial series truncated at $K=2$ and angular grid size of $N_\vartheta = N_\varphi = 30$.
	Similar results are achieved for angular grid size with incommensurable dimensions, e.g. $N_\vartheta = 29, N_\varphi = 31$.
	Solid line shows a cross-section of one exact corrugated flux surface.}
	\label{fig:tokamak_helical_pert_poincare}
\end{figure}

\noindent
In Fig.~\ref{fig:tokamak_helical_pert_poincare}, Poincar\'e plots of magnetic field lines obtained by the \red{quasi-}geometric
integration method for this perturbed configuration are shown at the cross section $\varphi=0$ together with a cross section
of one exact corrugated flux surface. It can be seen that the orbits from the \red{quasi-}geometric integrator become more chaotic with
increasing perturbation amplitude $\varepsilon_M$.

Diffusive behavior of orbits can be characterized by the variance of
the normalized toroidal flux $s$, accumulated over the time for the ensemble of test particles
starting from the same perturbed flux surface. This variance is described by the magnetic field line diffusion
coefficient $D_M^{ss}$ as $\left\langle \delta s^2\right\rangle=2 D_M^{ss} N$ where $N$ is the number of
toroidal orbit turns.
The effective diffusion coefficient $D_M^{ss}$ computed from the
orbits has a strong inverse scaling with poloidal $N_\vartheta$ and toroidal $N_\varphi$ grid
sizes and, furthermore, shows in general a small magnitude of diffusion even at coarse angular grid resolution.
E.g., at an angular grid size of $N_\vartheta = N_\varphi = 30$ the effective diffusion coefficient $D_M^{ss}$ is in the order of $10^{-13}$, $10^{-11}$ and $10^{-9}$
for relative perturbation amplitudes of $\varepsilon_M = 1~\%$, $3~\%$ and $10~\%$, respectively.
This level of field line diffusion is roughly five orders of magnitude smaller than observed for the initial version of the
code~\citep{eder_three-dimensional_2019}, and, in the worst case of $\varepsilon_M = 0.1$, results in stochastic diffusion
of electrons with the coefficient $D_\perp \sim D_M^{ss} v_\parallel r^2 / R \sim 100$ cm$^2$s$^{-1}$. For smaller $\epsilon_M$
values of the typical order for external magnetic perturbations in tokamaks this numerical diffusion is below the level of
classical electron diffusion and, therefore, can be safely ignored.

In \red{the }case of field aligned coordinates, chaotization of passing orbits (lines of force of the effective field $\bB^\ast$)
can only be caused by the cross-field drift.
Such a case is tested below for a strong violation of axial symmetry using as an example the stellarator field configuration
described in Ref.~\onlinecite{drevlak_quasi-isodynamic_2014}, namely,
a quasi-isodynamic reactor-scale device with five toroidal field periods and \red{a }major radius of 25 m.
Here the magnetic field has been normalized so that its modulus averaged over Boozer coordinate angles on the starting
surface is $B_{00}$~=~5~T.
Guiding-center orbits were computed with the \red{quasi-}geometric integration method in symmetry flux coordinates
for strongly passing electrons and ions with $v_\parallel / v = 0.9$  at the starting point on the flux surface $s = 0.6$ with an energy of 3~keV.
The numerical diffusion observed for a rather coarse angular grid with the size $N_\vartheta = N_\varphi = 30$
is roughly seven orders of
magnitude smaller than the minimum level of the neoclassical mono-energetic diffusion coefficient $D_{11}$
evaluated in section~\ref{sec:transport} for the same device with particles of the same energy.
\begin{figure}[h]
	\centerline{\includegraphics[keepaspectratio,width=1.2\linewidth]{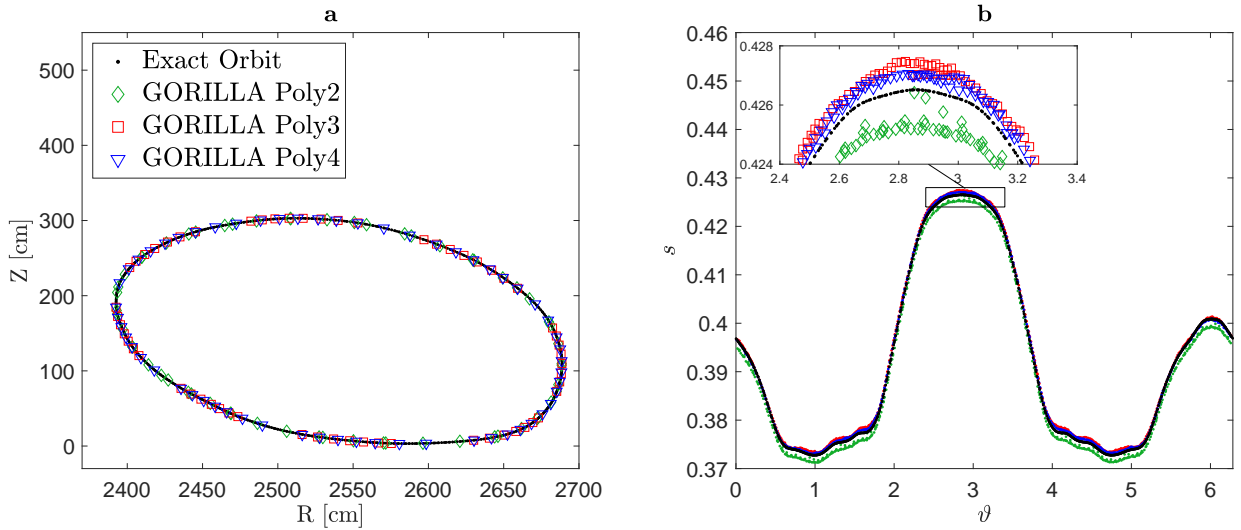}}
	\captionsetup{justification=raggedright,singlelinecheck=false,textfont=footnotesize,labelfont=footnotesize}
	\caption{(a \& b) Poloidal projection of Poincaré sections at $v_\parallel=0$ switching sign from $-$ to $+$ of a trapped 3~keV~D ion in 3D stellarator field configuration. Orbits evaluated by \bluex{\textit{Gorilla}}\redx{GORILLA} with the analytical solution in form of a polynomial series truncated at $K=2$ ($\Diamond$), $K=3$ ($\square$) and $K=4$ ($\triangledown$) are compared to the exact orbit ($\bullet$). Poincaré plot is depicted in cylindrical coordinates (a) and symmetry flux coordinates (b).\\
				\blue{(c) Evolution of the parallel adiabatic invariant $J_\parallel$ normalized to the value at $t=0$ in (a) and (b) over $10^5$~bounce times. Lines in (c) represent linear fits to the $J_\parallel$ data points: Exact result (solid) is compared to polynomial series truncated at $K=2$ (dash-dotted), $K=3$ (dotted) and $K=4$ (dashed).}}
	\label{fig:hydra_trapped_poincare}
\end{figure}

For the visualization of a trapped particle orbit, we use orbit footprints on Poincaré sections defined by the condition
$v_\parallel(\tau)~=~0$, i.e. phase space hypersurfaces containing orbit turning points.
\blue{Out of two kinds}\red{From the two types} of these surfaces, those are chosen \blue{where}\red{in which} the sign of $v_\parallel$ changes from negative to positive.
Fig.~\ref{fig:hydra_trapped_poincare} depicts a poloidal projection of orbit footprints corresponding
to a trapped ion with an energy of 3 keV. The orbits have been integrated in symmetry flux coordinates using the polynomial
series expansion in several orders and compared to the exact orbit computed with an adaptive RK4/5 integrator.
Further, the poloidal coordinates of the footprints have also been converted to cylindrical coordinates and visualized in
both coordinate systems, respectively.
Despite the same starting conditions \blue{of}\red{for} all orbits, the magnification in (b) clearly depicts \red{slight }differences in the shape of the orbits caused by a finite grid size ($100 \times 100 \times 100$) and different orders of the series expansion.

\begin{figure}[h]
	\centerline{\includegraphics[keepaspectratio,width=1.2\linewidth]{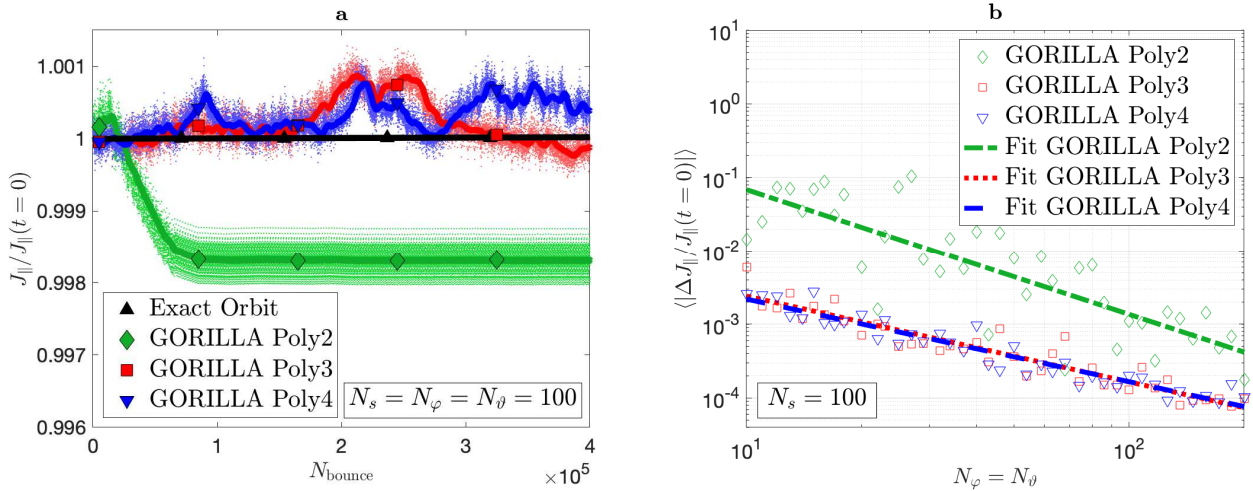}}
	\captionsetup{justification=raggedright,singlelinecheck=false,textfont=footnotesize,labelfont=footnotesize}
	\caption{\red{
				(a) Evolution of the parallel adiabatic invariant $J_\parallel$ normalized to the value at $t=0$ of the trapped particle orbit in Fig.~\ref{fig:hydra_trapped_poincare} over $4 \cdot 10^5$~bounce times. Lines in (a) represent window filtered trend of the $J_\parallel$ data points: Exact result ($\triangle$) is compared to polynomial series truncated at $K=2$ ($\Diamond$), $K=3$ ($\square$) and $K=4$ ($\triangledown$).\\
				(b) Modulus of the relative error of the parallel adiabatic invariant $J_\parallel$ averaged over $10^5$~bounce times as a function of the angular grid size $N_\varphi = N_\vartheta$. Orbits are evaluated by \bluex{\textit{Gorilla}}\redx{GORILLA} with the analytical solution in form of a polynomial series truncated at $K=2$ ($\Diamond$), $K=3$ ($\square$) and $K=4$ ($\triangledown$). The fits of the results are depicted with lines in accordance with the legend.}
				}
	\label{fig:hydra_trapped_parallel_invariant}
\end{figure}

In contrast to the axisymmetric tokamak field of the previous section, the parallel adiabatic invariant $J_\parallel$ is not
an exact invariant in a stellarator. Nevertheless, it should be well preserved as long as the trapped orbit stays within the same
class (traverses the same number of field minima over its bounce period) which is the case here.
Fig.~\ref{fig:hydra_trapped_parallel_invariant} \blue{(c)}\red{(a)} shows the time evolution of $J_\parallel$ for \blue{$10^5$}\red{$4 \cdot 10^5$} bounce periods of the corresponding orbits \red{of Fig.~\ref{fig:hydra_trapped_poincare}}. It can be seen that the results for the exact orbit shows no visible deviation for this configuration.
\red{The truncation of the polynomial series at $K=2$ causes a violation of the system's Hamiltonian structure.
This is manifested by an attractor in phase space, what is clearly visible in Fig.~\ref{fig:hydra_trapped_parallel_invariant} (a). A detailed analysis of the corresponding Poincaré sections when the attractor is already fully established (last $10000$ evaluations) reveals that the orbit strictly follows a continuous curve, staying in the same class.

However, truncation at higher orders ($K=3$ \& $K=4$) violates the Hamiltonian structure only negligibly and thus does not lead to non-Hamiltonian features. Nevertheless, in contrast to the exact orbit, $J_\parallel$ evaluated by \bluex{\textit{Gorilla}}\redx{GORILLA} with these polynomial orders is not accurately conserved. In particular, the value of $J_\parallel$ meanders randomly around the exact value which is caused by the diffusive behavior of the orbit induced by the piecewise linearization of the electromagnetic field.

Fig.~\ref{fig:hydra_trapped_parallel_invariant} (b) shows the modulus of the relative error of the parallel adiabatic invariant $J_\parallel$ averaged over $10^5$~bounce times as a function of the angular grid size $N_\varphi = N_\vartheta$ varied from $10$ to $200$.
At a moderate angular grid size of $N_\vartheta = N_\varphi = 28$ (in accordance with the Nyquist-Shannon sampling theorem~\cite{nyquist_certain_1928,shannon_mathematical_1948} explained below) this mean relative error stays already below $10^{-3}$ for the orders $K=3$ \& $K=4$.
Thus, the finite grid size used in the \red{quasi-}geometric orbit computation does not lead to a significant error accumulation.
With regard to the comparatively large number of bounce times, even the order $K = 2$ which is naturally the fastest with respect
to CPU time yields quite accurate results at sufficient grid resolution.
}

\blue{Furthermore, the finite grid size used in the geometric orbit computation does not lead to a significant error accumulation even for the order $K = 2$ which is naturally the fastest with respect
to CPU time.}
\section{Monte Carlo evaluation of neoclassical transport coefficients, performance benchmark}
\label{sec:transport}

Evaluation of neoclassical transport coefficients using the Monte Carlo
method~\cite{boozer_monte_1981,lotz_monte_1988} is widely
used for stellarators and tokamaks with 3D perturbations of the magnetic
field~\cite{wakasa_study_2008,tribaldos_monte_2001,isaev_venusf_2006,allmaier_variance_2008,velasco_calculation_2011,satake_neoclassical_2011,pfefferle_venus-levis_2014}.
An advantage of this method in its original, full-$f$ form is the use of test particle guiding-center
orbits without requiring \red{the }model simplifications needed in (more efficient) local approaches. \blue{Therefore}\red{The} Monte Carlo methods \red{thus }provide
an unbiased reference point in cases where those simplifications affect the transport \red{in a manner }such as \red{that }for regimes with
significant role of the tangential magnetic \red{drift~\cite{matsuoka_effects_2015,huang_benchmark_2017,velasco_knosos_2020}.}
An obvious disadvantage is that for realistic magnetic configurations Monte Carlo methods are CPU-intensive
with most of the CPU time spent for the integration of the guiding-center motion. The application of the proposed \red{quasi-}geometric integration method for this purpose instead of the usual Runge-Kutta method results in a visible speed-up
of the computations without significantly biasing the results.
Here, this application is made for benchmarking purposes \blue{assuming}\red{on the assumption} that the inaccuracies in orbit integration
which are tolerable in computations of transport coefficients are also tolerable
in global modelling of macroscopic plasma parameters.

The proposed orbit integration method is applied
within a standard Monte Carlo algorithm~\cite{boozer_monte_1981}
using the Lorentz collision model for the evaluation of the mono-energetic radial diffusion coefficient $D_{11}$.
The latter is determined via the average square deviation
of the normalized toroidal flux $s$ from its starting value $s_0$ as follows,
\be{eq:def_d11}
D_{11} = \frac{1}{2 t} \langle\left(s(t) - s_0 \right)^2\rangle.
\ee
Here, angle brackets $\langle\dots\rangle$ denote an ensemble average, $s(0)=s_0$, and the test particle tracing time $t$ is chosen to
be \blue{larger}\red{longer} than the local distribution function relaxation time $\tau_{\rm rel}$
\blue{and smaller}\red{but shorter} than the radial transport time, $t=10 \tau_{\rm rel}$.
A Monte Carlo collision operator identical to that of Ref.~\onlinecite{boozer_monte_1981} is applied here in-between
constant collisionless orbit integration steps $\Delta t$. These steps are small enough compared to the typical bounce time
$\tau_\mathrm{b}$ and collision time $\tau_\mathrm{c}$,
\be{eq:def_time_step_MC}
\Delta t = \min \left(\frac{\tau_\mathrm{b}}{20},\frac{\tau_{\mathrm{c}}}{20} \right).
\ee
Here, $\tau_{\mathrm{c}} = 1/\nu$ and $\tau_\mathrm{b} = 2 \pi R_0 / (v N_\mathrm{tor.})$ with $\nu$,  $R_0$, $v$ and $N_\mathrm{tor}$ denoting collisional deflection frequency, major radius, particle velocity and
number of toroidal field periods, respectively. The relaxation time $\tau_{\rm rel}$ is determined as the largest of
$\tau_{\mathrm{c}}$ and $\tau_\mathrm{b}^2/\tau_{\mathrm{c}}$.

\begin{figure}[t]
	\centerline{\includegraphics[keepaspectratio,width=1.1\linewidth]{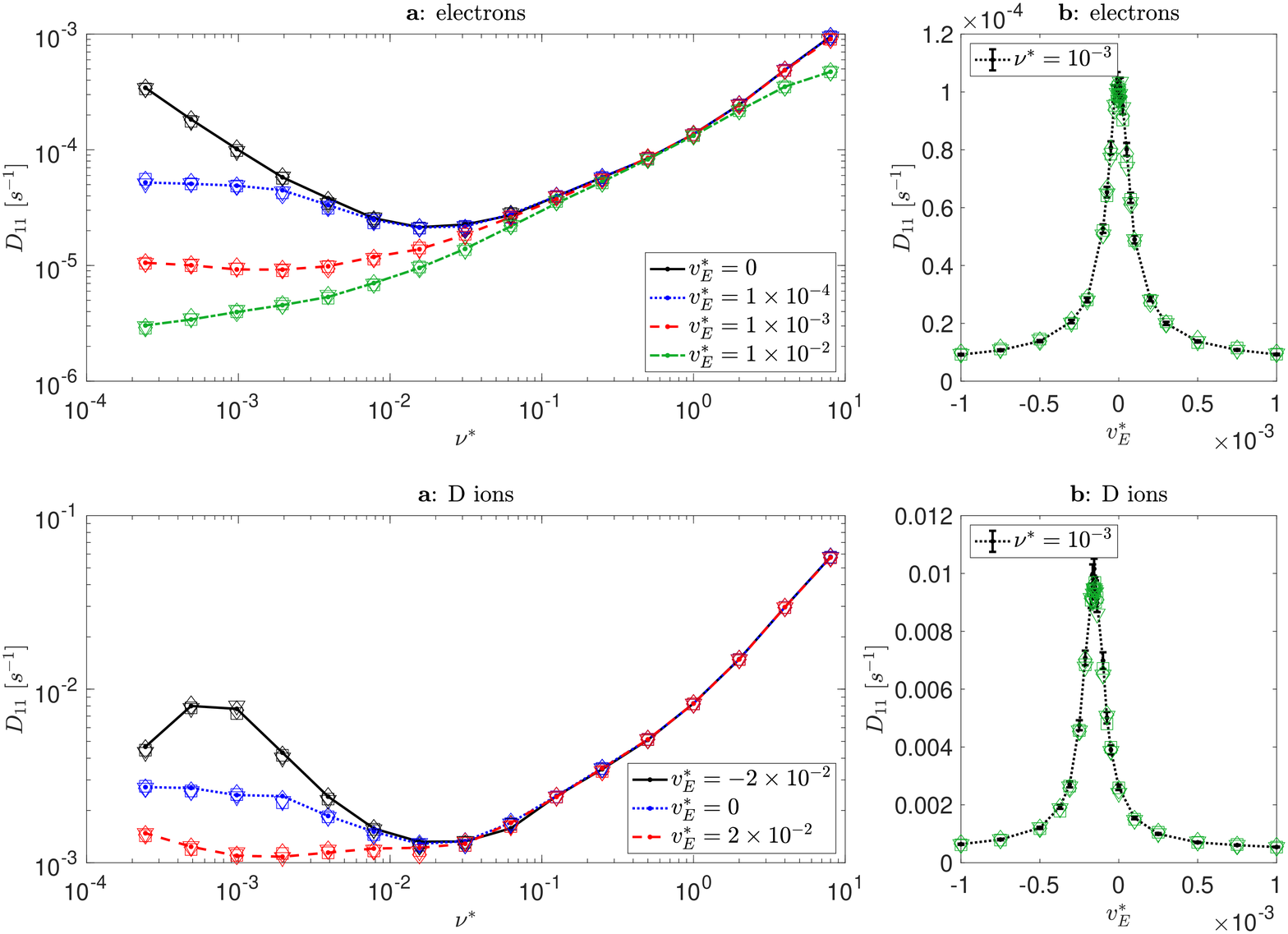}}
	\captionsetup{justification=raggedright,singlelinecheck=false,textfont=footnotesize,labelfont=footnotesize}
	\caption{Mono-energetic radial diffusion coefficients $D_{11}$ for electrons (top) and deuterium
	ions (bottom)
	as functions of (a) normalized collisionality $\nu^\ast$ and (b)  Mach number $v_E^\ast$.
	Lines of various styles (see the legends) - reference computation, markers - results of \red{quasi-}geometric
	integration with polynomial solution of the order $K$ for $K=2$ ($\Diamond$), $K=3$ ($\square$)
	and $K=4$ ($\triangledown$).
	Error bars indicate $95$~\% confidence interval.}
	\label{fig:transport_coefficient_electrons_dions}
\end{figure}
In the present example, the mono-energetic radial diffusion coefficient has been evaluated
for the quasi-isodynamic stellarator configuration~\cite{drevlak_quasi-isodynamic_2014}
used also for collisionless orbits in section~\ref{sec:orbits_3d}.
Guiding-center orbits were computed with the \red{quasi-}geometric integration method in symmetry flux coordinates
using polynomial series solutions of various orders $K$.
The grid size $N_s \times N_\vartheta \times N_\varphi = 100 \times 60 \times 60$ was selected to be appropriate \blue{to
minimize}\red{for minimizing} the numerical diffusion (see the previous section.) In a reference computation, guiding-center
equations~\eq{eqm_curv} in symmetry flux variables with electromagnetic field interpolated
by 3D qubic splines were integrated by an adaptive RK4/5 integrator.
\blue{In order to minimize statistical errors, computations have been performed for a large ensemble size
of $10000$ particles.}\red{Computations were performed for a large ensemble size of $10000$ particles in order to minimize statistical errors.}

The results for $D_{11}$ computed for 3 keV electrons and ions at $s_0=0.6$ are presented in
Fig.~\ref{fig:transport_coefficient_electrons_dions}.
Values of radial electric field $E_r$ and deflection frequency $\nu$, which determine
transport regimes, are respectively characterized here by two dimensionless
parameters~\cite{beidler_benchmarking_2011}, Mach number $v_E^\ast = c E_r /(vB_0 )$ and
collisionality $\nu^\ast = (R_0 \nu)/(\iota v)$, where $\iota$ is the rotational transform.
For the ions, in addition to the $\bE\times\bB$ rotation, also the tangential magnetic drift
plays a significant role which can be seen from the shift of the $D_{11}$ maximum on $v_E^\ast$
dependence. The results of \red{quasi-}geometric integration stay in agreement
with the reference computation within the 95 \% confidence interval in all cases even for the lowest
order polynomial solution $K=2$. \blue{Therefore, as shown below, a significant gain in the computation
time can be obtained in this kind of calculations.}\red{Therefore, in calculations of this kind a significant gain in the computation
time can be obtained as shown below.}

\begin{figure}[t]
	\centerline{\includegraphics[keepaspectratio,width=1.1\linewidth]{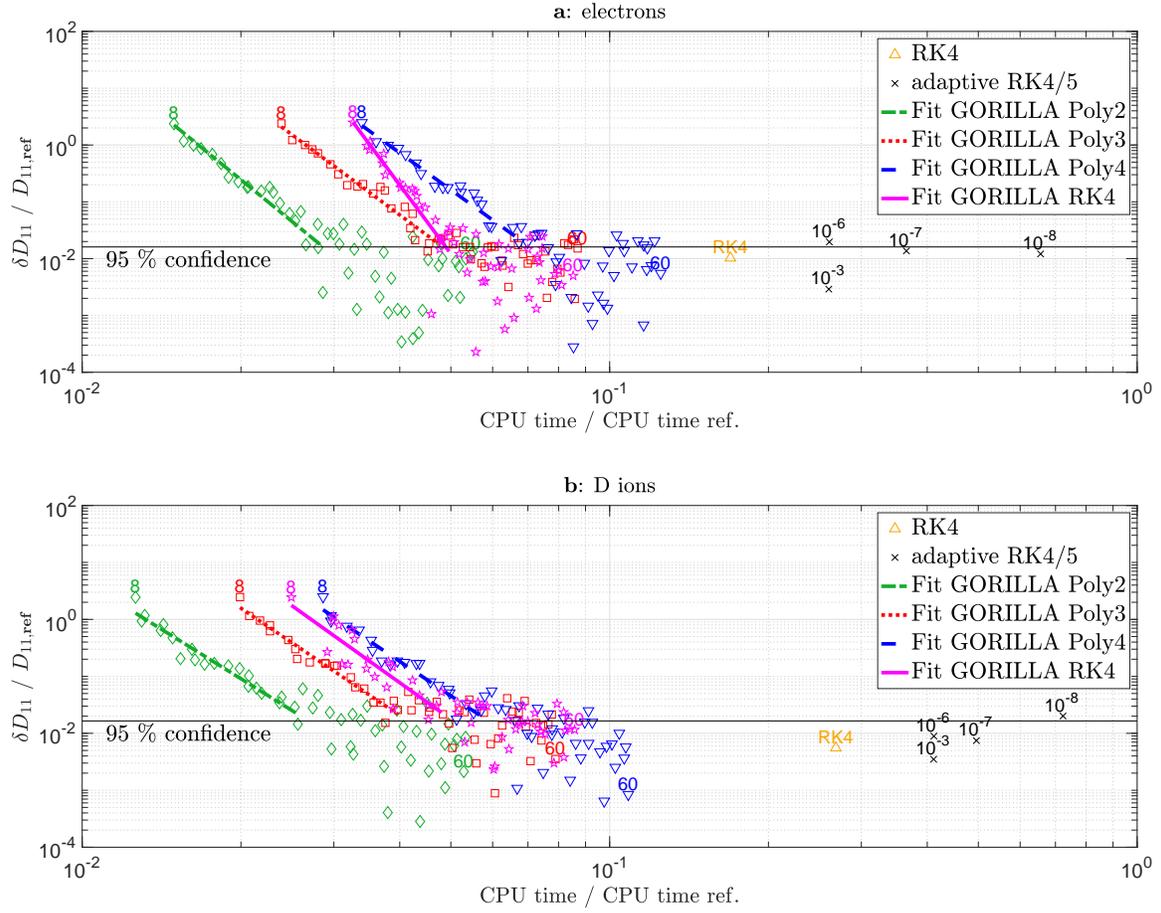}}
	\captionsetup{justification=raggedright,singlelinecheck=false,textfont=footnotesize,labelfont=footnotesize}
	\caption{Relative error of mono-energetic radial transport coefficient
	$D_{11}$ of electrons (top) and D~ions (bottom) vs. relative CPU time.
	\blue{Compared}\red{The compared} orbit integration methods are:
	Runge-Kutta 4 ($\star$),
	Adaptive RK4/5 with various relative errors indicated in the plot ($\times$),
	\red{quasi-}geometric integration with polynomial solution
	(\bluex{\textit{Gorilla Poly}}\redx{GORILLA Poly}) of the order $K=2$ ($\Diamond$), $K=3$ ($\square$) and $K=4$
	($\triangledown$), and with RK4 solution (\bluex{\textit{Gorilla RK4}}\redx{GORILLA RK4}, $\triangle$).
	\blue{Fits}\red{The fits} of \red{the }results are depicted with lines \blue{according to}\red{in accordance with} the legend.
	\blue{Random}\red{The random} error of the reference result, $D_{11,\rm{ref}}$,
	is depicted as a horizontal line limiting its 95~\% confidence interval.
	}
	\label{fig:cpu_speed_electrons_dions}
\end{figure}
Moreover, we compare the performance and scaling for parallel computation of guiding-center orbits
using the \red{quasi-}geometric orbit integration method with
computations using standard reference integrators (RK4 and adaptive RK4/5).
For this, different integrators have been used within $D_{11}$ computation described above
for a particular choice of dimensionless parameters, $v_E^\ast = 10^{-3}$ and $\nu^\ast =  10^{-3}$,
and an increased ensemble size of $30000$ test particles.
The numerical experiment has been performed on a single node of the COBRA cluster
of MPCDF with 40 CPU cores (Intel Xeon Gold 6126) running 80 concurrent threads with hyperthreading.

The reference value for the transport coefficient, $D_{11,\rm{ref}}$,
and the reference CPU time are obtained by orbit integration
with an adaptive RK4/5 integrator with a relative tolerance of $10^{-9}$.
The accuracy of the $D_{11}$ evaluation using different computation parameter settings
is represented by the relative error $\delta D_{11} / D_{11,\rm{ref}}$
where $\delta D_{11} = |D_{11} - D_{11,\rm{ref}}|$.
The CPU time purely used for orbit integration serves as
a measure for the computational effort of the methods.
This given CPU time does not contain any overhead operations, e.g. the construction of the grid, generation of random numbers for
pitch-angle scattering and computation of $D_{11}$ by evaluating
Eq.~\ref{eq:def_d11} with the help of a least-squares regression.

Fig.~\ref{fig:cpu_speed_electrons_dions} shows the relative error of the mono-energetic
radial transport coefficient versus the relative CPU time of computations using the
\red{quasi-}geometric orbit integration
method with the polynomial series solution of various orders, \bluex{\textit{Gorilla Poly}}\redx{GORILLA Poly},
and the iterative scheme with RK4 integration and Newton steps, \bluex{\textit{Gorilla RK4}}\redx{GORILLA RK4}.
Accuracy and CPU time of \red{quasi-}geometric orbit integrations have been varied by
mutually changing the angular grid size $N_\vartheta \times N_\varphi$
from $8 \times 8$ to $60 \times 60$ while keeping the radial grid size
constant at $N_s = 100$. In the stellarator configuration of
Ref.~\onlinecite{drevlak_quasi-isodynamic_2014} used here, the number of toroidal harmonic modes
per field period is $14$, leading to a minimum toroidal grid size $N_\varphi = 28$ in
order to satisfy the Nyquist-Shannon sampling theorem~\cite{nyquist_certain_1928,shannon_mathematical_1948}.
Therefore, regression lines are drawn for the range of data points with grid sizes
from $8 \times 8$ until $28 \times 28$, clearly showing a convergent behavior
of $D_{11}$ with increasing grid refinement.
Furthermore, the adaptive RK4/5 integration is additionally performed with
relative tolerances of $10^{-3}$, $10^{-6}$, $10^{-7}$ and $10^{-8}$, respectively.
Note that the computational
speed of the adaptive RK4/5 integration with a relative tolerance of $10^{-6}$ cannot be
increased by higher relative tolerances, e.g. $10^{-3}$, since the macroscopic Monte Carlo
time step, $\Delta t$ Eq.~\eq{eq:def_time_step_MC}, is already elapsed within a single
RK4/5 step with sufficient accuracy. Hence, also the non-adaptive Runge-Kutta~4 method
is tested, which naturally needs one field evaluation less per time step than RK4/5.
In all cases, the relative error of \red{the }RK4/5 and Runge-Kutta~4 results is determined here
mainly by statistical deviations, with a random error dominating the bias.

\blue{Besides}\red{Apart from} statistical errors due to Monte Carlo sampling, a limit for capturing
all toroidal and poloidal field harmonics is given by a minimum grid size of
two points per period due to the Nyquist–Shannon sampling theorem.
Fig.~\ref{fig:cpu_speed_electrons_dions}
visibly shows that statistical fluctuations already dominate the bias of all variants of the \red{quasi-}geometric
integration method above this sampling threshold, despite the large ensemble size of
30000 particles.
To avoid possible sampling artifacts at \red{an }even \blue{higher }higher particle count, we consider the \red{quasi-}geometric orbit integration method at the toroidal grid size $N_\varphi$ of
at minimum twice the number of toroidal modes in the magnetic field configurations.
The variant with the polynomial series solution truncated at $K=2$ (\bluex{\textit{Gorilla Poly 2}}\redx{GORILLA  Poly 2}) at
this grid resolution can be considered the fastest sufficiently accurate tested method to compute
$D_{11}$ for thermal ions and electrons. In \red{the }case of D ions with an energy of 3~keV this method
is one order of magnitude faster than the Runge-Kutta~4 integrator which is the fastest
reference method.

\section{Summary and outlook}
\label{sec:summary}

A \red{quasi-}geometric integration method for guiding-center orbits in general three-dimensional toroidal fields has been developed, implemented and presented here. This orbit integration procedure is based on a representation of the electromagnetic field by continuous piecewise linear functions using a spatial mesh.
Collisionless particle orbits in real space and magnetic coordinates and their respective invariants of motion have been studied in detail for axisymmetric and non-axisymmetric geometries. Due to the special formulation of the guiding-center equations, the magnetic moment and the total energy are conserved naturally. In \red{the }case of toroidal axisymmetry the canonical toroidal angular momentum is accurately preserved by the \red{quasi-}geometric method \redx{in third and fourth order series expansion
over the orbit parameter}, as well as the parallel \redx{adiabatic} invariant.
Thus, the property of such systems to ideally confine the orbits is retained.
\redx{In order to evaluate the limitations of these confinement properties, the kinetic energy of ions was chosen to be 300~keV. Otherwise systematic ODE integration errors originating from truncating the series expansion already at the second order would not be visible, since this error is proportional to the particle energy.}

For passing orbits in 3D fields, however, the piecewise linearization of the electromagnetic field
introduces some artificial chaotic diffusion which, nevertheless, could be made negligibly small by spatial grid refinement.
For trapped orbits in 3D fields, the approximate conservation of the parallel adiabatic invariant is not violated by significant
error accumulation.

To assess the method's performance, the mono-energetic radial transport coefficient, $D_{11}$, which gives a main contribution to neoclassical transport, has been evaluated for a quasi-isodynamic reactor-scale stellarator field~\cite{drevlak_quasi-isodynamic_2014} using the Monte Carlo method. For both, electrons and ions, the results obtained by \red{quasi-}geometric orbit integration are in good agreement with the results of adaptive RK4/5 integration with usual spline interpolation of electromagnetic fields.
In the performance benchmark, we observe that the \red{quasi-}geometric orbit integration method with the polynomial series solution truncated
at $K=2$ is the fastest sufficiently accurate tested method to compute $D_{11}$. For the case of D ions with an energy of
3~keV the guiding-center orbit integration is one order of magnitude faster than 4th order Runge-Kutta integration in splined fields.
\redx{Here, truncating the series expansion at the second order is not necessarily contradicting the result of collisionless guiding-center orbits, where a second order truncation leads to visible systematic errors at high kinetic energies. An appropriate choice of grid size and series expansion order strongly depends on the physical application, the kinetic particle energy and the complexity of the magnetic field (e.g. number of harmonic modes).}

For the application in global kinetic computations no extra effort is needed to obtain dwell times within spatial grid cells as
these are computed automatically in the present approach. Additionally, integrals of velocity powers over these dwell times
are available as analytical expressions.
The latter quantities are required for statistical scoring of orbits in Monte
Carlo computations of macroscopic parameters,
such as plasma response currents and charges caused by external non-axisymmetric
perturbations in tokamaks or parameters of the edge plasma in devices with 3D field geometry.
Moreover, similarly to the geometric integrator for axisymmetric two-dimensional fields described in Ref.~\onlinecite{kasilov_geometric_2016}, the presented method is less sensitive to noise
in the electromagnetic field than procedures relying upon high order polynomial interpolation.
These characteristics suggest additional overall performance enhancements in both numerical stability and computational efficiency, when the \red{quasi-}geometric orbit integration is applied to kinetic modeling.

\red{The applicability of the inherently low-order method to particle-in-cell
turbulence computations, where higher order schemes produce smoother solutions,
is still an open question.}
Finally, it should be mentioned that the method facilitates the coupling with kinetic neutral particle codes \blue{like}\red{such as}
EIRENE~\citep{reiter_eirene_2005}, where one needs to model particle conversion into neutrals and back with plasma
and neutral particles described in different coordinate systems. \blue{Necessary}\red{The necessary}
transformation of coordinates does not require solving any implicit dependencies (nonlinear equations),
since that is a linear operation in this approach and \red{is }therefore intrinsically fast.

\section*{Data Availability Statement}
The data that support the findings of this study are available from the corresponding author upon reasonable request.

\section*{Acknowledgements}

The authors would like to thank Michael Drevlak, Markus Meisterhofer, Artem Savchenko and Harry Mynick for respectively providing stellarator field configurations, Fortran advice, a tool for non-linear coordinate transformations and a tutorial for neoclassical transport in 3D systems, as well as Martin Heyn, Philipp Ulbl, Rico Buchholz, Patrick Lainer and Markus Richter for useful discussions. \red{Further thanks to three anonymous referees who have read the manuscript and suggested valuable improvements. }This work has been carried out within the framework of the EUROfusion Consortium and has received funding from the Euratom research and training programmes 2014–2018 and 2019–2020 under grant agreement no. 633053. The views and opinions expressed herein do not necessarily reflect those of the European Commission. The study was supported by the Reduced Complexity Models grant number ZT-I-0010 funded by the Helmholtz Association of German Research Centers. Support from NAWI Graz, and from the OeAD under the WTZ grant agreement with Ukraine No. UA 04/2017 is gratefully acknowledged.

\newpage

\appendix

\red{
\section{Hamiltonian structure of locally linear guiding-center equations}
\label{sec:appendix_hamiltonian}

\subsection{Liouville's theorem}
\label{sec:appendix_hamiltonian_Liouv}
Let us check that Liouville's theorem is fulfilled for the piecewise linear set represented locally by Eq.~\eq{standeqset} despite
the discontinuities of the phase space velocity at cell boundaries and the fact that the vector potential
is not exactly linked with magnetic field, i.e. that the relation
$\omega_c \left(\bB/\omega_c\right) = \nabla\times\bA$ is not fulfilled anymore
by the piecewise linear approximations $\omega_c^{(L)}$, $\left(B_i/\omega_c\right)^{(L)}$
and $A_i^{(L)}$ of the respective magnetic field parameters within the exact curvilinear coordinate metrics.
(In this section we use sub- or superscripts $(L)$ for quantities representing the
piecewise linear electromagnetic field.)
Formally, independent interpolation errors in these originally related
quantities result in a slight distortion of the metric tensor, but produce a
consistent set of equations of motion with a divergence-free modified magnetic field
$\mathbf{B}^*_{(L)}$ being a necessary condition for correct geometric properties.
Namely, $\mathbf{B}^*_{(L)}= \nabla \times \mathbf{A}^\star_{(L)}$ remains
divergence-free by using an interpolation of covariant components $A_k$
and covariant unit vector components $B_k/B \propto B_k/\omega_c$ that are
both given in a curl-compatible representation.
The gyrofrequency $\omega_c \propto B$ just plays the role of a scalar potential in Eq.~\eq{defU}
and doesn't affect the symplectic structure.

First, we notice (see Eq.~(44) of Ref.~\onlinecite{littlejohn_variational_1983}) that the
phase space Jacobian of the coordinate set $\by=(\bx,J_\perp,\phi,v_\parallel)$ where $\phi$ is the
%$(\bx,J_\perp,\phi,v_\parallel)$ where $\phi$ is the
gyrophase is
\be{Jacpp}
J=\difp{(\br,\bp)}{(\bx,J_\perp,\phi,v_\parallel)}=\frac{m_\alpha e_\alpha}{c}\sqrt{g}B_\parallel^\ast.
\ee
Liouville's theorem states that divergence of the phase space flow velocity is identically zero,
$J^{-1} \partial \left(J \dot{y}^i  \right) /{\partial y^i} \equiv 0$,
yielding
\be{liouv1}
\difp{}{x^i}J\dot x^i_{(L)} + \difp{}{J_\perp}J\dot J_\perp + \difp{}{\phi} J \dot\phi
+\difp{}{v_\parallel} J \dot v_\parallel^{(L)}=
\difp{}{x^i}J\dot x^i_{(L)}+\difp{}{v_\parallel} J \dot v_\parallel^{(L)} = 0,
\ee
where two terms vanished due to $\dot J_\perp=0$ and the independence of $\dot\phi$ of the gyrophase.
The remaining phase space velocity components are defined in accordance with~\eq{setofgceqs} and
a subsequent piecewise linear approximation as
\be{vels_L}
\dot x^i_{(L)} = \frac{1}{\sqrt{g}B_\parallel^\ast} \left(\frac{\rd x^i}{\rd\tau}\right)_{(L)},
\qquad
\dot v_\parallel^{(L)} = \frac{1}{\sqrt{g}B_\parallel^\ast} \left(\frac{\rd v_\parallel}{\rd\tau}\right)_{(L)}.
\ee
Here, $B_\parallel^\ast$ corresponds to the exact field while the derivatives with respect to $\tau$ are given
for the linearized field,
\bea{vels_tau}
\left(\frac{\rd x^i}{\rd\tau}\right)_{(L)}
&=&
\varepsilon^{ijk}\left(
v_\parallel \difp{A_k^{(L)}}{x^j}+v_\parallel^2\difp{}{x^j}\left(\frac{B_k}{\omega_c}\right)^{(L)}
+\left(\frac{B_k}{\omega_c}\right)^{(L)}\difp{U^{(L)}}{x^j}
\right),
\nonumber\\
\left(\frac{\rd v_\parallel}{\rd\tau}\right)_{(L)}
&=&
\varepsilon^{ijk}\difp{U^{(L)}}{x^i}
\left(
\difp{A_k^{(L)}}{x^j}+v_\parallel \difp{}{x^j}\left(\frac{B_k}{\omega_c}\right)^{(L)}
\right).
\eea
Note that here we had to replace the quantity $U^{(L)}$
(but not its spatial derivatives computed for constant $w$) back with its
expression~\eq{defU} via $v_\parallel$ (which remains exact also for the linearized field),
because $U^{(L)}$ is the only quantity containing the total energy $w$ which is not a constant parameter
but a function of phase space coordinates
for the derivatives in~\eq{liouv1}.
%Substitution of~\eq{vels_L}
%and then~\eq{vels_tau} in~\eq{liouv1} results in
Substitution of~\eq{vels_tau} in~\eq{vels_L}
and subsequently~\eq{vels_L} in~\eq{liouv1} yields

\bea{Oiszero}
&&\frac{m_\alpha e_\alpha}{c} \left( \difp{}{x^i}\left(\frac{\rd x^i}{\rd\tau}\right)_{(L)}
+\difp{}{v_\parallel}\left(\frac{\rd v_\parallel}{\rd\tau}\right)_{(L)} \right)
=\frac{m_\alpha e_\alpha}{c} \varepsilon^{ijk}\left(
v_\parallel \difp{^2A_k^{(L)}}{x^i\partial x^j}
+v_\parallel^2\difp{^2}{x^i\partial x^j}\left(\frac{B_k}{\omega_c}\right)^{(L)}
\right.
\nonumber
\\
&&+
\left.
\left(\frac{B_k}{\omega_c}\right)^{(L)}\difp{^2 U^{(L)}}{x^i\partial x^j}
+\difp{U^{(L)}}{x^j}\difp{}{x^i} \left(\frac{B_k}{\omega_c}\right)^{(L)}
+\difp{U^{(L)}}{x^i}\difp{}{x^j} \left(\frac{B_k}{\omega_c}\right)^{(L)}
\right)=0
\eea

which proves the theorem due to the symmetry of the expression in parentheses over $i$ and $j$.
Note that generally the second derivatives in~\eq{Oiszero} formally contain Dirac $\delta$ functions,
because the first derivatives of the piecewise linear functions are discontinuous at the cell boundaries.
However, the $\delta$ functions do not appear at a given boundary if one linearly transforms the spatial
variables $x^i$ so that one of the coordinate planes, e.g. $x^1=\mathrm{const.}$, contains the respective
tetrahedral cell face. This makes it evident that the normal component of the spatial velocity
$\left(\rd x^1/\rd \tau\right)_{(L)}$
is continuous at the cell boundary since it does not contain discontinuous derivatives over $x^1$
while the discontinuous tangential components
$\left(\rd x^{2,3}/\rd \tau\right)_{(L)}$ are not differentiated in~\eq{Oiszero} across the boundary
(over $x^1$).

\subsection{Symplecticity}
Let us show that the piecewise linearization of the electromagnetic field does not affect the
symplectic properties of the orbit geometry by using a similar to Ref.~\onlinecite{littlejohn_variational_1983} derivation of
the guiding-center equations.
First, we introduce the Lagrangian for the piecewise linear field,
\be{Lagr_L}
L^{(L)}=\frac{e_\alpha}{c}A_i^{\ast(L)}\dot x^i - J_\perp\dot\phi - H^{(L)},
\ee
where the independent phase space variables are $\by=(\bx,J_\perp,\phi,v_\parallel)$ and
\be{AstL}
A_i^{\ast(L)} = A_i^{(L)}+v_\parallel \left(\frac{B_i}{\omega_c}\right)^{(L)},
\qquad
H^{(L)}=\omega_c^{(L)} J_\perp+\frac{m_\alpha v_\parallel^2}{2} + e_\alpha \Phi^{(L)}.
\ee
The corresponding Euler-Lagrange equations,
\be{Eulag_gen}
\frac{\rd}{\rd t}\difp{L^{(L)}}{\dot y^i}=\difp{L^{(L)}}{y^i},
\ee
are explicitly given as
\bea{Eulag}
&&
\frac{e_\alpha}{c}\dot x^j\left(\difp{A_i^{\ast(L)}}{x^j}-\difp{A_j^{\ast(L)}}{x^i}\right)
+
\frac{e_\alpha}{c}\left(\frac{B_i}{\omega_c}\right)^{(L)}\dot v_\parallel
+\difp{}{x^i}\left(\omega_c^{(L)} J_\perp+ e_\alpha \Phi^{(L)}\right)=0,
\nonumber \\
&&
\dot\phi+\omega_c^{(L)}=0,
\qquad
\dot J_\perp=0,
\qquad
\frac{e_\alpha}{c} \left(\frac{B_i}{\omega_c}\right)^{(L)} \dot x^i-m_\alpha v_\parallel=0.
\eea
Making a convolution of the first vector equation of~\eq{Eulag} with the tensor
$\varepsilon^{ikl}\left(B_k/\omega_c\right)^{(L)}$ and using the last of Eqs.~\eq{Eulag}
one obtains an explicit expression for $\dot x^i$. The convolution of the same vector equation
with the vector $\varepsilon^{ikl}\partial A_l^{\ast(L)}/\partial x^k$ yields an expression
for $\dot v_\parallel$. All these phase space velocity components are expressed via~\eq{vels_tau}
as follows
\be{vels_LLag}
\dot x^i = \frac{1}{\left(\sqrt{g}B_\parallel^\ast\right)^{(L)}}
\left(\frac{\rd x^i}{\rd\tau}\right)_{(L)},
\qquad
\dot v_\parallel = \frac{1}{\left(\sqrt{g}B_\parallel^\ast\right)^{(L)}}
\left(\frac{\rd \bluex{x^i}\redx{v_\parallel}}{\rd\tau}\right)_{(L)},
\ee
and differ from~\eq{vels_L} by the first factor, where
\be{newjacL}
\left(\sqrt{g}B_\parallel^\ast\right)^{(L)} = \frac{e_\alpha}{m_\alpha c}\varepsilon^{ijk}
\left(\frac{B_i}{\omega_c}\right)^{(L)}\difp{A_k^{\ast(L)}}{x^j}.
\ee
By replacing in~\eq{Jacpp} the exact expression $\sqrt{g}B_\parallel^\ast$ with the linearized one of~\eq{newjacL}, we obtain
the Jacobian of the phase space coordinates $\by$ which formally have slightly modified
dependence on $(\br,\bp)$.

The preservation of the symplectic properties of the phase space flow by a piecewise lineariazation
of the field is obvious in case of 3D toroidal fields with embedded flux surfaces.
Using the canonical straight field line flux coordinate system~\cite{albert_symplectic_2020}
$\bx=(r,\vartheta,\varphi)$ where $A_r=B_r=0$ and, respectively, $A_r^\ast=0$ one can
introduce canonical momenta $\bP\equiv (P_1,P_2,P_3)=(P_\vartheta,P_\varphi,J_\perp)$
which are conjugates to the coordinates ${\bf Q}\equiv (Q^1,Q^2,Q^3)=(\vartheta,\varphi,-\phi)$
with
\be{canmom}
P_\vartheta=\frac{e_\alpha}{c}A_\vartheta^{\ast(L)},
\qquad
P_\varphi=\frac{e_\alpha}{c}A_\varphi^{\ast(L)}.
\ee
Then the Lagrangian~\eq{Lagr_L} and the Hamiltonian~\eq{AstL}
respectively take the form
\bea{canlagr}
L^{(L)} &=& P_i\dot Q^i-H^{(L)},
\qquad
\\
H^{(L)} &=& \omega_c^{(L)} J_\perp
+\frac{m_\alpha}{2} \left(\left(\frac{B_\varphi}{\omega_c}\right)^{(L)}\right)^{-2}
\left(\frac{c}{e_\alpha}P_\varphi-A_\varphi^{\ast(L)}\right)^2
+ e_\alpha \Phi^{(L)}.
\nonumber
\eea
Consequently,
the Euler-Lagrange equations~\eq{Eulag_gen} for the variable set $\by=({\bf Q},\bP)$
result in Hamilton's equations. Note that due to the continuity of the piecewise
linearization the mutual relations between canonical and non-canonical variables and
the Hamiltonian are continuous at the cell boundaries. Respectively continuous are
the orbits which fulfill the usual symplectic relations of mapping in time.

Note that both, Eqs.~\eq{vels_L} and Eqs.~\eq{vels_LLag} result in the same phase
space orbit geometry with the latter having Hamiltonian time dynamics described by
\be{timedyn}
\frac{\rd t}{\rd \tau}=\left(\sqrt{g}B_\parallel^\ast\right)^{(L)}.
\ee
\bluex{These}\redx{High accuracy of this} dynamics, however, \bluex{are}\redx{is} not important for the steady state problems which are here of main interest.
Since none of the coefficients of the kinetic equation depend
in this case on time variable, it can be replaced by $\tau$ resulting in
\be{kineq}
\difp{f}{\tau}+\left\{f,H\right\}_\tau=-\difp{}{y^i}J F_C^i(f)  + Jq,
\ee
where $\left\{a,b\right\}_\tau$ are modified Poisson brackets~\eq{poisson}, $J$ is the
Jacobian~\eq{Jacpp}, $F_C^i(f)$ is phase space flux density due to collisions,
and $q$ is some phase space particle source.
Obviously, the subsequent linearization of the fields does not violate
the conservation properties of Eq.~\eq{kineq}. Moreover, Boltzmann's distribution
$f=f_B(H)$ remains to be a steady state solution since it commutes with the
Hamiltonian in the Vlasov part and results in $F_C^i(f_B)=0$ for the background in the thermodynamic equilibrium.
}
\newpage

\section{Analytical integrals of velocity powers over the dwell time}
\label{sec:appendix_integrals}

In this section it is presented how analytical expressions for the dwell time integrals of $v_\perp^2$, $v_\parallel$ and $v_\parallel^2$ can be obtained. Furthermore, the latter quantity is exemplary given as an explicit expression.

We start with the exact polynomial series solution of ODE set~\eq{standeqset}
\be{eq_appendix_repeat}
\bz=\bz_0 + \sum\limits_{k=1}^\infty \frac{\tau^k}{k!} \left(\hat \ba^{k-1}\cdot\bb + \hat \ba^k\cdot \bz_0\right),
\ee
already given in Eq.~\eq{explser}. For the parallel velocity $v_\parallel = z^4$ this series can be written in form of a shifted exponential function
\be{vpar_exp}
v_\parallel (\tau) = e^{\alpha \tau} \left( v_{\parallel,0} + \frac{\beta}{\alpha} \right) - \frac{\beta}{\alpha},
\ee
where $\alpha$, $\beta$ and $v_{\parallel,0}$ stand for the matrix element $a_{44}$, the vector component $b^4$ and the initial value for the parallel velocity at the cell entry, respectively. After squaring Eq.~\eq{vpar_exp}, one can perform a Taylor series expansion of the orbit parameter up to the $4^{\text{th}}$ order,
\bea{}
v_\parallel^2 (\tau) &\approx& v_{\parallel,0}^2 + \tau (2 \beta v_{\parallel,0} + 2 \alpha v_{\parallel,0}^2)
+ \tau^2 \left( \beta^2 + 3 \alpha \beta v_{\parallel,0} + 2 \alpha^2 v_{\parallel,0}^2 \right)\\ \nonumber
&+& \tau^3 \left( \alpha \beta^2 + \frac{7}{3} \alpha^2 \beta v_{\parallel,0} + \frac{4}{3} \alpha^3 v_{\parallel,0}^2 \right)
+ \frac{1}{12} \tau^4 \left( 7 \alpha^2 \beta^2 + 15 \alpha^3 \beta v_{\parallel,0} + 8 \alpha^4 v_{\parallel,0}^2 \right).
\eea
This is the highest order, where an algebraic expression of the dwell time $t_d$ to pass a cell can be found.
In order to obtain the dwell time integral of $v_\parallel^2$, its polynomial representation can simply be integrated
\bea{}
\int_0^{t_d} v_\parallel^2 (t) \mathrm{d}t &\approx& C  \left(\tau_d v_{\parallel,0}^2
+ \frac{1}{2} \tau_d^2 (2 \beta v_{\parallel,0} + 2 \alpha v_{\parallel,0}^2)
+ \frac{1}{3} \tau_d^3 \left( \beta^2 + 3 \alpha \beta v_{\parallel,0}
+ 2 \alpha^2 v_{\parallel,0}^2 \right) \right.
%\nonumber
\\
&+& \frac{1}{4} \tau_d^4 \left( \alpha \beta^2 + \frac{7}{3} \alpha^2 \beta v_{\parallel,0}
+ \frac{4}{3} \alpha^3 v_{\parallel,0}^2 \right)
%\nonumber \\ &+&
+
\left. \frac{1}{60} \tau_d^5 \left( 7 \alpha^2 \beta^2
+ 15 \alpha^3 \beta v_{\parallel,0} + 8 \alpha^4 v_{\parallel,0}^2 \right) \right),
\nonumber
\eea
where $C=\rd t /\rd \tau = B_\parallel^\ast\sqrt{g}=\mathrm{const.}$ within the cell.
Here, this lowest order approximation of $C$ is intended, because for the moments of a steady state 
\eq{kineq} only the integrals over the orbit parameter $\tau$ are needed 
\redx{to be computed accurately. The reason is that ther error of such an approximation is always
small as long as the field quantities are well resolved by the grid. In turn, the cross-field drift terms 
which formally scale with the Larmor radius must be
accounted accurately in $\tau_d$ as well as the respective integrals because they determine the orbit dynamics near the banana tips. In the case that higher
order accuracy of the moments with respect to the grid size is required the
derivative $\rd t / \rd \tau$ must be used in its form~\eq{timedyn}. Due to Eq.~\eq{newjacL} this 
derivative is a product of a linear function of the coordinates and a linear function of $v_\parallel$.
Therefore, the dwell time $t_d$ and the time integrals can still be computed analytically
leading to somewhat more complicated expressions}. 
\red{Note that here the dwell time $\tau_d$ may be a sum of entry and exit times in the case of a pre-defined time step.}

Clearly, the dwell time integral of $v_\parallel$ requires to omit squaring of Eq.~\eq{vpar_exp} and to proceed straightforwardly in the same manner.

Moreover, the squared perpendicular velocity $v_\perp^2$ is purely a function of position inside a tetrahedral cell due to its proportionality with the cyclotron frequency, $v_\perp^2 = 2 \omega_c \frac{J_\perp}{m}$, which is a piecewise linear function of position in the geometric orbit integration formulation. Thus, the dwell time integral of $v_\perp^2$ is obtained via an integral along the orbit which can easily be computed by using its polynomial representation given in Eq.~\eq{eq_appendix_repeat}.

\newpage

\bibliography{gorilla_paper}{}
\end{document}